\documentclass[aps,prd,twocolumn,nofootinbib,superscriptaddress]{revtex4-1}
\usepackage{amsmath} % Useful for equation alignment
\usepackage{verbatim} % Allows to have comments over multiple lines
\usepackage{mathrsfs}
\usepackage{amsfonts}
\usepackage{hyperref}

\def\l{\langle}
\def\r{\rangle}

\def\l{\langle}
\def\r{\rangle}

\newcommand{\py}{{\hat{\pi}}}     % its conjugate momentum
\newcommand{\pyRI}{\hat{\pi}^{R,I}}

\newcommand{\fluc}[1]{(\Delta #1)^2}      % square of fluctuations
 
\newcommand{\cre}{\hat{a}^{\dagger}}    % creation

\newcommand{\ann}{\hat{a}}            % and annihilation operators

\newcommand{\tc}{\eta_{\vec{k}}^c}
\newcommand{\beq}{\begin{equation}}
\newcommand{\eeq}{\end{equation}}
\newcommand{\nk}{\vec{k}}

\newcommand{\x}{\vec{x}}
\newcommand{\y}{\vec{y}}
\newcommand{\bra}{\langle}
\newcommand{\ket}{\rangle}
\newcommand{\bea}{\begin{eqnarray}}
\newcommand{\eea}{\end{eqnarray}}

\begin{document}

\author{Susana Landau}
\email{slandau@df.uba.ar}
\affiliation{Instituto de F\'isica de Buenos Aires, Ciudad Universitaria - Pab. 1, 1428 Buenos Aires, Argentina}

\author{Gabriel Le\'on}
\email{gabriel.leon@nucleares.unam.mx}
\affiliation{Dipartimento di Fisica, Universit\`a di Trieste, Strada Costiera 11, 34151 Trieste, Italy.}
\affiliation{Instituto de Ciencias Nucleares, Universidad Nacional Aut\'onoma de M\'exico, M\'exico D.F. 04510, M\'exico}

\author{Daniel Sudarsky}
\email{sudarsky@nucleares.unam.mx}
\affiliation{Instituto de Astronom\'ia   y F\'isica  del Espacio, Casilla de Correos  67,  Sucursal 28,  1428 Buenos Aires, Argentina}
\affiliation{Instituto de Ciencias Nucleares, Universidad Nacional Aut\'onoma de M\'exico, M\'exico D.F. 04510, M\'exico}

\title{Quantum origin of the primordial fluctuation spectrum and its statistics}

\begin{abstract}
The usual  account  for the origin of cosmic structure  during inflation is not  fully  satisfactory,  as it   lacks  a physical mechanism
capable of   generating
   the inhomogeneity and anisotropy of our Universe,  from an exactly homogeneous and isotropic initial state associated
   with the early inflationary regime.
 The  proposal  in  [A. Perez, H. Sahlmann, and D. Sudarsky, Classical Quantum Gravity,
  \textbf{23}, 2317, (2006)]  considers  the spontaneous  dynamical collapse of the wave function,   as  a possible answer to that  problem.
  In this  work,  we review  briefly the difficulties   facing the standard approach,  as  well  as the  answers  provided  by the above  proposal   and  explore their  relevance
    to the investigations  concerning the  characterization of the primordial spectrum and  other   statistical   aspects  of the cosmic microwave background  and  large-scale matter distribution.
     We  will see that the  new approach  leads to  novel ways of  considering some of the relevant  questions,  and, in particular, to distinct  characterizations
     of the non-Gaussianities  that might  have left imprints  on the  available  data.
\end{abstract}


%%%%%

\maketitle
\section{Introduction}
\label{Intro}

At what point in the cosmic evolution do the actual
primordial inhomogeneities arise? In other words, when
does our Universe depart from the exceedingly high homogeneity and isotropy\footnote{The  level of  inhomogeneity  that might  still be present at  any point   during inflation is  expected to  be  of order  $e^{-N}$,
   where $N$ is the  number of e-folds  of inflation occurred up to that point.} that is thought to result from the first
stages of inflation? This is a question that one might expect
should be addressed, at least, in principle, by any theory that
deals with the emergence of cosmic structure. Yet in the
standard inflationary account \cite{1}, which is nowadays regarded as a remarkable success, the context in which such
issues can be addressed seems to be simply absent \cite{2}. That
is, within the orthodox accounts, one cannot identify the
physical process responsible for the generation of those
features in our Universe. In fact, according to the inflationary paradigm, from a relatively wide initial set of possibilities marking the end of the mysterious quantum gravity era, the accelerated inflationary burst leads to a homogeneous and isotropic (H\&I) Universe where the quantum
fields are all characterized by the equally homogeneous
and isotropic vacuum states (usually taken specifically to
be the so-called Bunch-Davies vacuum). From these conditions, it is usually argued, in a rather unclear\footnote{Acknowledgments that this  is
 an  unclear  aspect  of  the  standard approach can be   seen,   for instance,  in   the   book   \textit{Cosmology}  by Weinberg \cite{3}, where  the author explicitly states his view on the subject.} although
strongly image-evoking manner, that the ``quantum fluctuations'' present in such a quantum state morph into the seeds of anisotropies and inhomogeneities that characterize our
late Universe. This issue is sometimes characterized as the
``transition from the quantum regime to the classical regime,'' but we find this a bit misleading: most people would
agree that there exist no distinct and separated classical and
quantum regimes. The fundamental description ought to be
always quantum mechanical; the so-called \emph{classical regimes}
are those in which certain quantities can be described to a
sufficient accuracy by their classical counterparts representing the corresponding quantum expectation values. The
paradigmatic example of this classical regime is provided
by the coherent states of a harmonic oscillator, which correspond to minimal wave packets with expectation values of
position and momentum that follow the classical equations
of motion (Ehrenfest theorem). In any case, it seems clear
that from a situation corresponding to a H\&I background,
and quantum fields characterized by a H\&I state, one cannot
end up--in the absence of something else, which in other
circumstances would be identified as a measurement, but
which clearly cannot be invoked in the present setting\footnote{Observers  and measuring apparatuses  are only possible  well after the  H\&I has been broken,  so those  can hardly  be  part of the cause of the breakdown.}--in a
situation that is characterized, at any level, as containing
actual inhomogeneities and anisotropies. It is clear that, in
terms of the standard dynamics, such a transition cannot
be accounted for by anything that relies just on the
gravity/inflaton action,\footnote{In fact,   even the interaction with other fields,    controlled  by   the  usual symmetry preserving  dynamics,  cannot account for the  emergence of   inhomogeneities and   anisotropies,  since,    according  to the inflationary  paradigm,  the state  for all fields should correspond to a homogenous and isotropic  state  such as the  Bunch-Davies vacuum. } which is known to preserve such symmetries. Simply put, if the initial state is H\&I and the
Schr\"odinger evolution is tied to a Hamiltonian that preserves
these symmetries, the resulting state cannot be anything but
a H\&I state (see Appendix \ref{BD}). Nonetheless, various types of
arguments are often put forward in attempts to bypass the
above conclusion. Most cosmologists adopt a posture that
attributes to decoherence the role of explaining the emergence of inhomogeneities and anisotropies from the H\&I
state. This approach faces several problems:

\begin{enumerate}
 \item  The decoherence program is based on partitioning
the degrees of freedom in two categories--The degrees of freedom of the environment and the degrees
of freedom of the system of interest. In the cosmological case however, in which one cannot evoke observers or measurements, the way to do the separation of the degrees of freedom is rather \emph{ad hoc}.

\item In order to argue that the symmetry was broken,
one needs to assume that the world is suddenly
represented by one of the elements appearing in
the diagonal of the deciphering density matrix,
and it is not clear how to argue for that.
 
\item Sometimes people evoke the many worlds interpretation of quantum theory in order to deal with the
previous point but seem to ignore that, in order to
do that, one needs to choose a privileged basis
associated with the world splittings, and that choice,
in practice, is tied to the notion of conscience, again
a notion that cannot be invoked in the context at
hand. Another popular posture is to rely on the
consistent histories approach, ignoring the problematic issues afflicting that proposal. In particular, we
should note that the usage of the formalism requires
a choice of realm, a choice that in the current context
seems completely arbitrary. In fact, one can make
one such choice when one is led to the conclusion
that the Universe is, even today, perfectly homogenous and isotropic (see Appendix \ref{Problems}).

\end{enumerate}

The extended discussion of the conceptual problems
inherent to quantum theory and those associated to its
application to the cosmological situation at hand have
been presented in previous works by some of us and by
others in Refs. \cite{4,5,6}. The main message is that the problem
we face is tied with the so-called measurement problem of
quantum theory and that this problem becomes exacerbated in the present case, in which we are dealing with
cosmology, a field in which the standard ways to address
such problems are simply unavailable \cite{7}. In this work, we
reproduce all those arguments in detail, mentioning them
only briefly, as the main objective of the present manuscript is to focus in the statistical aspects that emerge in
this context (a slightly more detailed discussion of those is
offered in Appendix \ref{Problems} for the benefit of the reader).

We will discuss a new way of looking at those issues,
based on what we consider to be a conceptually more
transparent picture that relies on a modified version of
the standard inflationary paradigm, which we have been
advocating in previous works \cite{4,8,9,10,11}. The basis of that
proposal is to modify the standard inflationary paradigm
with the incision of a modified quantum mechanics that
involves the spontaneous collapse of the wave function.

We should note, however, that we cannot escape from
the related problems, even if we choose to adopt a very
``pragmatic position'': Assume, that one chooses to ignore
the shortcomings of the standard accounts \cite{5} and accepts
that, say, decoherence addresses somehow the issue at hand
and that the mystery lies ``only'' in the question concerning
the precise mechanism that lies behind the fact that, from
the options exhibited in those analyses (i.e., the options
displayed in the diagonal reduced-density matrix; see
Appendix \ref{Problems}), one single realization seems to be selected
\cite{12} for our Universe. In adopting such a point of view, one
would be assuming that the initial symmetry has been lost
(at least for practical purposes) in association with that
particular ``realization'' or ``actualization'' (represented by
a particular element in the density matrix). Thus, it seems
clear that, for the sake of self-consistency, one should
consider, when studying aspects of the inhomogeneity
and anisotropies in the cosmic microwave background
(CMB) we observe, the state corresponding to such realization or actualization, and not the complete vacuum state,
which describes the H\&I state of affairs previous to the
actualization.\footnote{The reliance on a particular realization or actualization refers,
of course, to the fact that, according to the standard arguments,
the resulting density matrix, after becoming essentially diagonal
due to decoherence, would be taken to represent an ensemble of
universes, with our particular one corresponding to one of the
elements occurring in the diagonal of that matrix. That state
should then be considered as somehow ``selected by nature'' to
become realized (or to be the one we perceive). If one wanted to
consider the issue at a deeper level, one would have to face the
question of what such actualization represents at the theoretical
level and what is, if any, the physics that controls it.
Alternatively, one might take the view (often referred as the
many worlds interpretation) that these other universes are somehow also realized, and thus they exist in realms completely
inaccessible to us. In that case, the actualization corresponds
to that Universe in which we happen to exist.} In following such views, the discussion that
we are presenting in this paper would have to be taken to
represent the effective description corresponding to ``our
perceived Universe'' (in a context in which one puts
together something like the many worlds interpretation,
with the arguments based on decoherence). Although we
definitely do not adhere to such a view for the reasons
explained in Ref. \cite{5}, it is clear that when accepting a
description, such as the one presented above, one would
have to use the characteristics of the selected state in order
to estimate the details of the inhomogeneities and anisotropies in the cosmic structure and its imprints in the CMB.

As we indicated, the purpose of this paper is to discuss
the manner in which the consideration of statistical aspects
of the CMB and the large-scale matter distributions should
be modified when taking into account the modifications
needed to explain the emergence of inhomogeneities and
anisotropies in terms of theories incorporating something
like the spontaneous collapse of the wave function. The
need to rely on a different approach to study things like
the non-Gaussianities in the CMB, arises, in part, due to the
vastly larger space of possibilities for exotic effects, which
opens in connection to the unknown dynamics of the
collapse processes. In other words, in the standard treatments, the spectrum would be determined by the inflationary
theory (number of fields, kinetic terms, and interacting
potentials), and the nature of the initial state, while in the
approach we have been advocating a novel source of statistical anomalies, is provided by the details of the modification of quantum theory by the dynamics of collapse.

One example of these novel possibilities is provided by
the study of the details of the mode by mode collapse
within the semiclassical treatment of the problem as developed in Ref. \cite{13}. In that work, it was found the collapse 
of a mode with comoving wave vector $\vec{k}_0$ must be tied with
the modification of the state of the field in the higher
harmonics of that mode. It was found, in particular, that
the effect would be stronger for mode $2 \vec{k}_0$. This, in turn,
leads to the consideration of the possibility of strong
correlations in the collapse parameters of the two modes,
an effect that would produce a particular type of exotic
correlations--it is unclear if they should be called
non-Gaussianities as they involve modifications of the
two-point functions--something that would produce a
particular kind of signature in the CMB \cite{14}.

The organization of this manuscript is as follows: In
Sec. \ref{stand1}, we offer a preliminary discussion of the posture we
advocate regarding the emergence of structure and its
implications for the statistical analysis of the CMB and
some aspects of the usual approach focusing on the aspects
we consider to be conceptually unclear. In Sec. \ref{stand2}, we
review the standard picture for primordial non-Gaussianities. In Sec. \ref{colapso}, we review the collapse models
description for the inflationary origin of the seeds of the
cosmic structure. In Sec. \ref{further}, we focus on the statistical
aspects as seen from our perspective of the primordial
inhomogeneities, propose new characterizations of the
non-Gaussianities, and discuss new measures to be associated with the bispectrum. Finally, in Sec. \ref{disc}, we discuss
our findings. We use conventions in which the signature
of the space-time metric is $(-,+,+,+)$ and units where
$c=1$ but will keep the gravitational constant $G$ and $\hbar$
appearing explicitly throughout the paper.

\section{Some preliminaries on the  emergence of features of  the CMB  and  statistical considerations}
\label{stand1}

Let us start this section by noting that, in the usual
accounts, it is hard to pinpoint where exactly the statistical
aspects enter at the theoretical level, how that is connected
to the kind of statistics one considers at the observational
level, and which kind of statistics one is dealing with. That
is, in the standard approach, our specific Universe is not
described in any sense (not even in terms of unknown yet
explicitly identified quantities), and the randomness that
one invokes, as characterizing the relation of theory and
observation, lies hidden in unspecified aspects associated
with the vagueness of the interpretations. In other words,
one cannot identify the random variables; one does not
know how many there are, and one cannot say how exactly
the various elements of the ensemble of Universes differ
from each other. One imagines an ensemble of universes
and assumes that their collective departure from H\&I is
somehow characterized by the H\&I vacuum state or the
state that results form the unitary evolution thereof (despite
such a state being homogeneous and isotropic). One then
considers that the ensemble is being described, while each
of the individual elements of the ensemble cannot be
described, or that its description is irrelevant.

Within such setting, one proceeds to make, either explicitly or implicitly, the assumption that statistics over
such an ensemble correspond to the statistics, over time,
over space, or over orientation, in our particular Universe.
In fact, one assumes that they are all equivalent. It should
be clear that such assumptions are, therefore, taken to say
something about the individual element of the ensemble,
and it is not completely clear what it is. If our Universe is
not described by the quantum state we use in our equations,
what can we say about it? In order to look for justification
and clarification of such identifications, we must turn to the
quantum theory from which one expects to extract the
predictions. The problem is that, while quantum theory
has a clear and workable interpretation (even if not completely satisfactory \cite{15}) for dealing with laboratory experiments (the Copenhagen interpretation), for which the
measuring devices and observers can be taken as clearly
identified, for the case of the cosmological problem at
hand, we are faced with a situation deprived of such
entities that normally provide an interpretation.

Thus, the issue we will be addressing cannot be turned
into one of ``measurement,'' while implicitly assuming that
such concept can be used in the delicate quantum mechanical context examined in this paper. This is simply because
as we have already noted, cosmology needs to account for
the emergence of the conditions\footnote{Primordial inhomogeneities are supposed to evolve into galaxies and galaxy clusters, and within galaxies, stars and planets
are supposed to arise by gravitational collapse, and then life is
supposed to arise in the appropriate circumstances on some
planets, particularly on Earth.} that make such things as
observers and apparatuses possible to start with.

In order to fully and satisfactorily address the problem at
hand, it seems we need to be able to point out ``what
exactly is wrong with the argument leading to the conclusion drawn above. In other words, where does nature deviate from the theory leading to the erroneous conclusion that our Universe is, even today, at the fundamental
quantum level, perfectly homogeneous and isotropic?'' It
follows that such explanation must indicate where the
ordinary $U$ evolution--with the symmetry preserving
Hamiltonian--breaks down.

We can easily see that none of the proposals to deal with
the issue, and which are based on the standard paradigms,
can single out any point where that breakdown might occur
or, much less, point to a physical reason for that departure
from standard quantum theory (we turn the interested
reader toward Appendix \ref{div}, where we explore in more
detail these issues and justify more precisely our point of
view).

This has led us to take a view that ties this problem with
the ideas advocated by L. Diosi and R. Penrose, which
argue \cite{16,17} that quantum theory should itself suffer
modifications as a result of its combination with the fundamental theory of space-time structure.\footnote{This is what is often thought of as quantum gravity. We did
not use that term because that often presupposes that one is
considering the relevant theory to be simply the adaptation of
general relativity to the standard quantum theory, while what one
has in mind, when following Diosi and Penrose’s ideas, is
something much more distant from known physics, involving,
as indicated, modifications of quantum theory itself.} Among the aspects of the theory that would be substantially affected
according to those views are those related to the reduction postulate (or $R$ process) and its contrast with the unitary
evolution (or $U$ process) controlled by Schr\"odinger's equation. In fact, the issue of dynamical quantum reduction has
received a lot of attention within the community working
in foundational aspects of quantum theory, and there are, in
the existing literature, several rather well-defined proposals in this regard, such as those in Refs. \cite{15,16,18,19,20}.
The proposal behind our work is based on the hypothesis
that a dynamical collapse of the wave function lies behind
the breakdown of the initial homogeneity and isotropy. In
other words, a nonunitary ``jump'' in the quantum state of
the system plays a role in transforming the inflaton vacuum
into a quantum state that lacks the translational and rotational symmetries of the former state.
It goes without saying that we cannot, at this stage, try or
hope to point out the precise physical origin of such
dynamical collapse.\footnote{In particular, collapse theories are known to face, in principle,
serious difficulties with Lorentz and general covariance and
issues related to conservation laws. However, important advances have been made in addressing both classes of issues
(Refs. \cite{21,22}), even if we cannot say we have at our disposal
anything resembling a completely satisfactory theory.} However, once one has accepted
that something of this sort is occurring, one can parametrize its basic characteristics and use the relevant observational data to infer something about the nature of the novel
physics that must lie behind such phenomena. This has
been the basic attitude behind the program started in Ref. \cite{4}. We should emphasize, that although most of our
work has centered on that rather simplistic collapse model
developed specifically for the cosmological problem at
hand, the discussion of most of this paper would apply
equally to more general models and, in particular, to
approaches based on exciting proposals like Ghirardi-Rimini-Weber \cite{18} and continuous spontaneous localization \cite{23}. In fact, some recent works are devoted to the adaptation of the continuous spontaneous localization theory for its application to the problem of the emergence of
inhomogeneities and anisotropies in cosmology \cite{24,25}.

Here, we want to focus on the impact of such ideas on
the statistical study of the CMB. We will discuss some
delicate interpretational aspects related to quantum theory,
its implicit usage in the standard approach to the study of
the CMB, and its characterization in terms of a spectra as
well as the accounts of the origin of cosmic structure. We
will briefly explore here, for the first time, some of the
basic differences associated with statistical considerations,
between those tied to the usual approach and those approriate to our proposal.

In order to make things a bit more explicit, let us start by
reminding the reader that in the standard approaches, the
study of the statistical nature of the problem is based on the
study of the statistical $n$-point functions of the Newtonian
potential, $\overline{\Psi(x_1) \ldots \Psi(x_n)}$, with the overline denoting the
average over an ensemble of universes. Having no access
to such ensemble, one needs to face the issue of what
the relationship between those $n$-point functions and the
quantities we actually measure is. Moreover, one needs
to consider how these quantities are connected with the
quantum $n$-point functions. The usual approach \cite{26,27}
relies on the identification

\beq
 \overline{\Psi (x_1)\ldots\Psi(x_n)}= \langle 0 | \hat{\Psi} (x_1) \ldots \hat{\Psi} (x_n) | 0 \rangle,
\eeq
where $\langle 0 | \hat{\Psi} (x_1) \ldots \hat{\Psi} (x_n) | 0 \rangle$
 is a standard quantum mechanical $n$-point function for the quantum field operators
(corresponding to the vacuum state at hand). That is, one is
making the identification of quantum and statistical $n$-point
functions. As we said, the latter are naturally associated
with an ensemble of universes, all of which, even if real,
are unaccessible to us. The usual line of argument continues by invoking ergodic arguments, to make a further
connection between ensemble averages and time averages,
with other vague arguments indicating one might replace
the latter with spatial averages and often turning, in practice, to orientation averages. On the other hand, at the
quantum level, the interpretation is, as we noted before,
even more problematic. In the standard laboratory situations, one has an apparatus designed to measure a certain
observable $\mathcal{O}$, and quantum theory then indicates that, in
each individual measurement, one would obtain an eigenvalue of the corresponding operator. Furthermore, immediately after the measurement, the individual system is taken to be in the state corresponding to the resulting eigenvalue (as immediate repetition of the same
measurement in such an individual system gives, with
probability 1, the exact same value). The sudden change
in the state of the system is known as the state function
reduction or wave function collapse and is thought as being
brought up by the measurement (in fact, the interpretation
is not fully satisfactory, but we have become used to the
fact that in laboratory situations it works). Moreover, the
quantum expectation value of the observable $\bra \xi | \mathcal{O} | \xi \ket$ in
the state $|\xi \ket$ (the system's state before the measurement)
should be equal to the average of the observed values of the
corresponding quantity, over a large enough ensemble of
identical systems. It is important to note here that such an
interpretational scheme works only as long as a clearly
identified measurement is involved, as one essential aspect
of the nature of the quantum world is that one cannot
consistently adopt a point of view advocating that the
measurement simply served to reveal a preexisting value
of such a quantity (see, for instance, Ref. \cite{28}).

Let us illustrate these and other related issues by
considering the simplest place where one can appreciate
the problematic aspects of such identifications: the case of
the one-point function. Let us focus here on the standard
treatment that relies on the so-called Mukhanov-Sasaki
variables, defined by

\begin{equation}\label{defs}
u\equiv \frac{a\Psi}{4\pi G \dot{\phi}_0},\quad v\equiv a\left(\delta\phi+\frac{\dot{\phi}_0}{\mathcal{H}}\Psi\right),
\end{equation}
where  $ \Psi $ is  the  metric  perturbation known as  the Newtonian potential, $ \dot{\phi}_0$ is the  derivative of the background inflaton
with respect to conformal time $\eta$, $ \delta\phi$ is the perturbation in  the inflaton  field, $ a$ is the  scale factor,
and ${\mathcal{H}} \equiv \frac{\dot{a}}{a}$ (related to the standard Hubble parameter $H$ through $\mathcal{H}=aH$).
 Einstein's equations then lead to
$\Delta u=z\left(\frac{v}{z}\right)^{\cdot}  $
and
$  v=\frac{1}{z}\left(zu\right)^{\cdot}$
where $z\equiv\frac{a\dot{\phi}_0}{\mathcal{H}}$.
 Given the  equations of motion, the Newtonian potential can  thus be expressed  in
  terms of the field  $v(\x,\eta)$ and its momentum  conjugate $\pi_v (\x,\eta) = \dot{v} (\x, \eta)$.
  The expression for the   corresponding Fourier components  is

\begin{equation}\label{source.Psi}
{\Psi}_{\vec{k}} (\eta) =-\frac{\sqrt{4\pi G\epsilon}H}{k^2}\left({\pi}_{v \vec{k}} (\eta) -\frac{\dot{z}}{z}{v}_{\vec{k}} (\eta) \right),
\end{equation}
where $\epsilon$ is the so-called slow-roll parameter $\epsilon \equiv 1-\dot{\mathcal{H}}/\mathcal{H}^2$.

We are interested  in the temperature anisotropies of the CMB observed today on the celestial
two-sphere, which  are related  to the inhomogeneities in the Newtonian potential on the last scattering surface,
\begin{equation}\label{deltaT}
\frac{\delta T}{T_0} (\theta,\varphi) = \frac{1}{3} \Psi (\eta_D, \vec{x}_D).
\end{equation}

The  data are  described  in terms of the coefficients  $\alpha_{lm}$
of the multipolar series expansion
\begin{equation}\label{expansion.alpha}
\begin{split}
\frac{\delta T}{T_0}(\theta,\varphi)=\sum_{lm}\alpha_{lm}Y_{lm}(\theta,\varphi),\\
\alpha_{lm}=\int \frac{\delta T}{T_0}(\theta,\varphi)Y^*_{lm}(\theta,\varphi)d\Omega,
\end{split}
\end{equation}
here $\theta$ and $\varphi$ are the coordinates on the celestial two-sphere, with $Y_{lm}(\theta,\varphi)$ as the spherical harmonics.

 The  value  for the  quantities $\alpha_{lm}$  are then given by

\begin{equation}\label{alm2}
\alpha_{lm} = \frac{4 \pi i^l}{3}   \int \frac{d^3{k}}{(2 \pi)^3} j_l (kR_D) Y_{lm}^* (\hat{k}) \Delta (k)
%(\eta_R,\eta_D)
 \Psi_{\vec{k}} (\eta_R),
\end{equation}
with $j_l (kR_D)$ as the spherical Bessel function of order $l$; $\eta_D$
is the conformal time of reheating, which can be associated
with the end of the inflationary regime, and $R_D$ is the
comoving radius of the last scattering surface. We have
explicitly included the modifications associated with latetime physics encoded in the transfer functions $\Delta (k)$.

Now, the problem is that, if we compute the expectation
value of the right-hand side (i.e., identifying $\langle \hat{\Psi} \rangle = \Psi$) in
the vacuum state $| 0  \rangle$, we obtain 0, while it is clear that for
any given $l, m$, the measured value of this quantity is not 0.\footnote{ We are ignoring the remote possibility that, just by coincidence, and for some specific $l$ and $m$, the quantity $\alpha_{lm}$ would
vanish within the observational margin of error. As can be seen
in Sec. \ref{colapso}, according to our point of view, that would require a
remarkable cancelation between terms determined by a large
collection of random numbers.} That is, if we rely in this case on the one-point function and
the standard identification, we find a large conflict between
expectation and observation. We might even be tempted to
say that evidence of non-Gaussianity has already been
observed in each measurement of a particular $\alpha_{lm}$ . This
is, of course, not what one wants. Advocates of the standard approach would indicate that $ \langle \alpha_{lm} \rangle =0 $ is not to be
taken as ``the prediction of the approach'' regarding our
Universe and that this would only hold for an ensemble of
universes. The issue, of course, is what precise interpretational posture regarding the theory can be used to justify
this, while at the same time justifying the positions taken \emph{vis-\`a-vis}  the other quantities that emerge from the theory
(such as the higher $n$-point functions). A theory that depends on a case by case adaptation of an interpretational
rule is not a very satisfactory theory. However, this makes
clear that disentangling the various statistical aspects
(ensemble statistics; space and time statistics, including
orientation statistics; and, finally, the nature of the assumed
connection of quantum and statistical aspects) and making
explicit the assumptions underlying the identifications, as
well as the expected limitations, is paramount to avoid
confusion and to allow the judging of a theory on its true
merits.

As a matter of fact, it seems clear that anything that can
be considered as a satisfactory approach should enable
one to understand what exactly is wrong with the above
argument. First, let us note that, just as the Fourier
transform of a function is a weighted average (with weight
$e^{i\nk \cdot \x}$), so are the spherical harmonic transforms of functions. Thus, $\alpha_{lm}$ is a weighted average over the last
scattering surface (cosmologists often refer to the average
over the sky) because it is an integral over the celestial
two-sphere of $\frac{\delta T}{T}$ 
 weighted with a given function, the $Y_{lm}$.
The common argument in the literature, as we have noted,
indicates that averaging over the sky justifies the identification of observations with quantum expectation values.

In other words, the argument indicates that the relevant
prediction (obtained in terms of quantum expectation values) concerns the ensemble averages, and these should be
equal to spatial averages and the latter to averages over the
sky. However, apparently, this should not hold for weighted
averages over the sky (otherwise, all the $a_{lm}$'s would be 0).
If not, why not? There seems to be no clear answer.

Namely, if we take the theoretical estimate as

\begin{eqnarray}\label{almR}
\alpha_{lm}^{\textrm{th}} = \frac{4 \pi i^l}{3}   \int \frac{d^3{k}}{(2 \pi)^3} j_l (kR_D) Y_{lm}^* (\hat{k}) \Delta (k)
\langle 0 | &\hat{\Psi}_{\vec{k}}& (\eta_R)| 0\rangle \nonumber \\
 &=&0,
\end{eqnarray}
and compare it with the measured quantity $\alpha_{lm}^{\textrm{obs}}$, we would
find a large discrepancy. The answer, within the standard
accounts, would need to be that, for some reason, in order
to be allowed to make identifications, we should invoke a
further averaging: the average over orientations. Only then
would we have any confidence that our estimates are
reliable. Now, let us ask ourselves the question of why
this should be; it seems completely unclear. Anyhow, the
point is that we would be asked to compute

\begin{equation}\label{almR Av}
 \overline{ \alpha_{l}} = \frac{1}{2 l +1}  \sum_m \alpha_{lm},
\end{equation}
and we would then expect this quantity to be zero.

We need to confront the following issues:

\begin{enumerate}
 \item   Why is that so? Why can this average be expected to
yield zero but not each individual $\alpha_{lm}$ as in Eq. \eqref{almR}?

\item  Empirically, does this hold? In other words, is the
actual average of observed complex quantities in
Eq. \eqref{almR Av}, in fact, zero, or is it not?

\end{enumerate}

Regarding the first question, it seems imperative to
choose a suitable interpretational framework in order to
be able to decide \emph{a priori} what the appropriate identifications are and also to be able to evaluate whether or not we have a good theoretical understanding. It appears that, in the standard way of looking at the issue, there is really no
justification to expect anything but the vanishing of each
$\alpha_{lm}$. We must avoid getting confused with the notion that quantum theory involves uncertain predictions. The point
is that the only part of quantum theory that involves such
indeterminism is the measurement process, and we do not
want to call upon that in this particular situation. It is
true that, even in ordinary laboratory situations, the
``measurement problem'' is quite unsettling. However, in
the case at hand, the problem is exacerbated because we
cannot even contemplate any physical observer or measuring device existing prior to the emergence of the seeds of
structure. Thus, we cannot even rely on our old battle tool:
the Copenhagen interpretation, which explains the non-vanishing of those quantities that predates both the growth
of galaxies and the existence of ``observers and measuring
devices.''

Regarding the second issue, we would like to comment
the work of Armendariz-Picon \cite{29}, which starts to address
(albeit in a rather limited way, because the analysis is done
for a very small number of values of $l$) that question. The
results of this work indicate that the $\overline{\alpha_{l}}$ are small (one order of
magnitude smaller than the variance of the $\alpha_{lm}$ , that is, $\sqrt{C_l}$), and that seems reassuring. But is this sufficient? Is
that what we should expect according to our theory? Why?
Should it not be zero up to the actual experimental errors\footnote{Here, we should be careful in considering the sources of
error: As in any measurement, we have the systematic errors and
the statistical errors associated with uncontrolled disturbances but we should not confuse statistics over several determinations
of a specific $\alpha_{lm}$, say, with different experimental runs or with
different satellites, and the statistics for a fixed $l$ over the
orientation number $m$. For a fixed value of $l$, the variability of $\alpha_{lm}$  with $m$ should not, in our view, be taken as some statistical
error but as truly valuable data containing valuable information
about the physics behind the emergence of the seeds of cosmic
structure.}
in the observations? Evidently, these are just rhetorical
questions, raised only to show that it is easy to be confused
regarding the comparisons of theory and observations, if
one accepts, without questioning, the usual arguments
given by the standard approach. It seems evident that, in
order to have a clear answer to those questions, one needs
to have a precise and unambiguous characterization of
what exactly the mapping between the theory and the
measured quantities is. Actually, it seems one would
need to consider those comparisons as tests of whether
the identifications one is making are or are not appropriate
ones.

It is our view, as advocated in Refs. \cite{4,5}, that the
standard paradigm has no satisfactory answers to these
issues. We hope this brief discussion serves to illustrate
the problem we must face concerning the identification of
theoretical predictions and observations in the situation at
hand.

We end this section by reminding the reader that, if one
wants to consider the average value of any quantity, it is
imperative to specify over which set the average is defined.
There are just no ``averages'' as absolute concepts. In the
remainder of the manuscript, we will make an important
differentiation between averages over ensembles of universes, averages over a spacelike hypersurface, averages
over the last scattering surface, and averages over orientations. The question we want to address is how we are able to compare the theoretical estimates, based on quantum
expectation values for some quantities, with measured
values of related quantities. The approach we will be
primarily focusing on is the one pioneered in Ref. \cite{4}
and which seems to have more potential for dealing univocally with such questions than the standard one.

\section{The standard picture for the primordial non-Gaussianities}
\label{stand2}

This section will briefly review the standard accounts on
the primordial non-Gaussianities following closely
Refs. \cite{30,31,32}. There is absolutely no original work in
this section or any extensive discussion of our views
(just a few relevant comments); we simply present here
the usual treatment on the subject following what is
commonly found in the literature, in order to compare it
with our own approach, and discuss the main differences.
For more details and derivations, we refer the reader to the
comprehensive review by Komatsu \cite{33}, Bartolo \textit{et al}.
\cite{34}, and the references cited therein. 

Historically, non-Gaussianity, as a test of the accuracy of
perturbation theory, was first suggested by Allen \textit{et al}. \cite{35}.
However, most of its importance to date relies on the
premise that it will play a leading role in furthering our
understanding of two fundamental aspects of cosmology
and astrophysics \cite{36}:

\begin{enumerate}
 \item  the physics of the very early Universe that created
the primordial seeds for large-scale structures,

\item  the subsequent growth of structures via gravitational
instability and gas physics at later times.

\end{enumerate}

Within the standard approach, by non-Gaussianity, one
refers to any small deviations in the observed fluctuations
from the random field of linear, Gaussian, curvature perturbations. The curvature perturbations, $\Psi$, generate the CMB anisotropy, $\delta T/T$. The linear perturbation theory gives a linear relation between $\Psi$ and $\delta T/T$ on large scales (where the Sachs-Wolfe effect dominates) at the decoupling epoch, i.e., $\delta T/T \sim (1/3) \Psi$. It follows from the relation, $\delta T \propto \Psi$, that  if $\Psi$  is Gaussian, then $\delta T$ is
Gaussian, but what exactly does one mean by Gaussian
at the observational level?

One of the most important results of the inflationary
paradigm is that the CMB anisotropy arises due to
curvature perturbations, which, in turn, are produced
by quantum fluctuations. In the standard single-field
slow-roll scenario, these fluctuations are due to fluctuations of the inflaton field itself, when it slowly rolls
down its potential $V(\phi)$. Within this approach, the primordial perturbation is Gaussian; in other words, its
Fourier components are uncorrelated and have random
phases. When inflation ends, the inflaton $\phi$ oscillates
about the minimum of its potential and decays, thereby
reheating the Universe.

In the inflationary paradigm, the perturbations of the
field $\delta \phi$ and the perturbations of the curvature $\Psi$ are treated as standard quantum fields, \footnote{In fact, they are both part of a unified field $v$.} evolving in a classical quasi-de Sitter background space-time. The quantity of observational interest is called the power spectrum of the curvature perturbation $P_{\Psi} (k,\eta)$. The power spectrum is obtained from

\beq\label{2ptosvac}
\langle 0 | \hat{\Psi} (\x, \eta) \hat{\Psi} (\y, \eta) | 0 \rangle,
\eeq
where $|0 \rangle$ is called the Bunch-Davies  vacuum and represents the initial state of the field $\hat{v}$, which is the Mukhanov-Sasaki field variable defined in \eqref{defs} (for a discussion about the symmetric properties of the Bunch-Davies vacuum, see Appendix \ref{BD}).

It is precisely at this step where a subtle issue arises, namely that, in the standard picture, one is given various and distinct arguments (e.g. decoherence, horizon crossing, many worlds interpretation of quantum mechanics, etc.) to accept the identification

\beq\label{qmclas}
\langle 0 | \hat{\Psi} (\x, \eta) \hat{\Psi} (\y, \eta) | 0 \rangle = \overline{\Psi (\x, \eta) \Psi (\y, \eta)},
\eeq
where $\Psi(\x,\eta)$ now stands as a classical stochastic field and the overline denotes the average over an ensemble of Universes. In other words, the value of the field $\Psi$ in each point $(\x,\eta)$ varies from each one of the members of the ensemble of ``Universes,'' with a variance $\overline{\Psi^2}$. Therefore, the power spectrum $P_\Psi(k,\eta)$ is defined in terms of the Fourier components of $\Psi(\x,\eta)$ by

\beq
\overline{\Psi_{\nk} (\eta) \Psi_{\nk'} (\eta) } \equiv (2\pi)^3 \delta (\nk+\nk') P_\Psi (k,\eta).
\eeq
Consequently, the power spectrum is related to the two-point function through

\beq
\overline{\Psi (\x, \eta) \Psi (\y, \eta)} = \int_0^\infty \frac{dk}{k} \mathcal{P}_\Psi (k,\eta) \frac{\sin kr}{kr},
\eeq
with $r\equiv |\x - \y|$, and we also used the definition of the \emph{dimensionless} power spectrum $\mathcal{P}_\Psi (k,\eta) \equiv P_\Psi (k,\eta) k^3 / 2\pi^2$. The variance $\overline{\Psi^2}$ is given by

\beq\label{varpsi1}
\overline{\Psi^2 (\x, \eta)} = \int_0^\infty \frac{dk}{k} \mathcal{P}_\Psi (k,\eta).
\eeq

The expression \eqref{varpsi1} diverges generically. In particular, we know that the spectrum of the primordial curvature
perturbation is roughly $P_\Psi (k,\eta) \propto k^{-3}$. That is, $\mathcal{P}_\Psi (k,\eta)$ is nearly constant (i.e. independent of $k$); therefore, Eq. \eqref{varpsi1} diverges in a
 logarithmic way for $k\to 0$ and $k\to \infty$. The way the standard pictures deal with this issue \cite{27} is to establish a $k_\textrm{max}$ equal to the ``horizon,'' and work in a cubic box of physical size $aL$ much larger than the Hubble radius. Thus,

\beq\label{varpsi2}
\overline{\Psi^2 (\x, \eta)} \simeq \mathcal{P}_\Psi (\eta)  \int_{L^{-1}}^{aH} \frac{dk}{k} =  \mathcal{P}_\Psi (\eta) \ln \frac{aL}{H^{-1}}.
\eeq
That is, in order to avoid the divergence in $\overline{\Psi^2}$, one is forced to introduce some particular values of $k$ as cutoffs (for a detailed discussion related to this fact, see Appendix \ref{div}).

The question that arises now is how can we evaluate any  average over an ensemble of Universes if we have observational access to just one--our own--Universe.  The  obvious answer is  that we cannot.
Normally, one is presented  with  ergodic  arguments indicating that averages over time  should be   equated  with  ensemble  averages. However, ergodicity  relies on   equilibrium and   the inflationary regime
is not one of equilibrium. Furthermore,  ignoring that issue, we  would  need to  find  an  argument justifying the identification  of time  averages and spatial averages, presumably to be  carried over the
hypersurface corresponding to the time of decoupling. Then, we need to  make  sure our argument  applies only to direct averages and not to weighted  averages,  as we discussed in the introduction. And
finally,  as  we  do not have access (at least using the CMB)  to that  whole  hypersurface, nor to  any large open region within it,  but only to the portion of it that   intersects our past light cone (the two-sphere
known as the last scattering surface), we  must  find  some argument indicating we can replace such spatial averages   with averages over orientations.

In the next section, we will show how our approach  deals with these questions. In the remainder of this  section, we will accept the validity of Eq. \eqref{qmclas} and ignore those  issues.

If $\Psi(\x, \eta)$ is Gaussian\footnote{That is, there exists some physical mechanism for which the quantum variable $\hat{\Psi}(\x, \eta)$ becomes a classical stochastic field $\Psi(\x,\eta)$ with Gaussian distribution.}, then the two-point correlation function \eqref{2ptosvac} specifies all the statistical properties of $\Psi(\x, \eta)$, for the two-point correlation function is the only parameter in a Gaussian distribution. If it is not Gaussian, then we need higher-order correlation functions to determine the statistical properties.

For instance, a nonvanishing three-point function\footnote{Just as  in the case of  the two-point correlation function, the standard approach relies on the identification
\beq
\langle 0 | \hat{\Psi} (\x, \eta) \hat{\Psi} (\y, \eta)  \hat{\Psi} (\vec{z}, \eta) | 0 \rangle = \overline{ \Psi (\x, \eta) \Psi (\y, \eta)  \Psi (\vec{z}, \eta) }.
\eeq
}

\beq
\overline{ \Psi (\x, \eta) \Psi (\y, \eta)  \Psi (\vec{z}, \eta) }
\eeq
is an indicator of non-Gaussian features in the cosmological perturbations. The Fourier transform of the three-point function is called the \emph{bispectrum}\footnote{In the following, we will not write the explicit dependance of the conformal time $\eta$ unless it leads to possible confusion.} and is defined as

\beq\label{bispec}
\overline{ \Psi_{\nk_1} \Psi_{\nk_2} \Psi_{\nk_3}} \equiv (2\pi)^3 \delta (\nk_1 + \nk_2 + \nk_3 )B_\Psi (k_1, k_2,k_3).
\eeq
The importance of the bispectrum comes from the fact that it represents the lowest-order statistics able to distinguish non-Gaussian from Gaussian perturbations.

The delta function in Eq. \eqref{bispec} enforces the triangle condition, that is, the constraint that the wave vectors in Fourier space must close to form a closed triangle, i.e. $\nk_1+\nk_2+\nk_3=0$. Different inflationary models predict maximal signal for different triangle configurations. The standard approach of the study of the structure of the bispectrum is usually done by plotting the magnitude of $B_\Psi (\nk_1,\nk_2,\nk_3) (k_2/k_1)^2 (k_3/k_1)^2$ (with $|\vec{k}_i| \equiv k_i$) as a function of $k_2/k_1$ and $k_3/k_1$ for a given $k_1$, with a condition that $k_1 \geq k_2 \geq k_3$ is satisfied. The usual classification of various shapes of the triangles uses the following names: squeezed $(k_1 \simeq k_2 \gg k_3)$, elongated $(k_1 = k_2 + k_3)$, folded $(k_1 = 2k_2 = 2k_3)$, isosceles $(k_2=k_3)$ and equilateral $(k_1 = k_2 = k_3)$. Within the cosmology community \cite{37,38,39}, these shapes of non-Gaussianity are potentially a powerful probe of the mechanism that creates the 
primordial perturbations.

One of the first (and most popular) ways  to parametrize non-Gaussianity phenomenologically was via a small nonlinear correction to the linear Gaussian perturbation  \cite{40,41},

\begin{eqnarray}\label{psipert}
\Psi(\x,\eta)  &=& \Psi_\textrm{L} (\x,\eta) + \Psi_\textrm{NL} (\x,\eta) \nonumber \\
&\equiv& \Psi_\textrm{L} (\x,\eta) + f_{\textrm{NL}}^{\textrm{loc}} [ \Psi_\textrm{L}^2 (\x,\eta) - \overline{\Psi_\textrm{L}^2 (\x,\eta)} ],
\end{eqnarray}
where $\Psi_\textrm{L} (\x,\eta)$ denotes a linear Gaussian part of the perturbation, and the variance $\overline{\Psi_\textrm{L}^2 (\x,\eta)}$ is implemented in the same sense as presented in Eq. \eqref{varpsi2}. Henceforth, $f_{\textrm{NL}}^{\textrm{loc}}$ is called the \emph{local nonlinear coupling parameter} and determines the ``strength'' of the primordial non-Gaussianity. This parametrization of non-Gaussianity is local in real space and therefore is called \emph{local non-Gaussianity}. In this \emph{local model}, the contributions from ``squeezed'' triangles are dominant, that is, with, e.g., $k_3 \ll k_1, k_2$.  Using Eqs. \eqref{psipert} and \eqref{bispec}, the bispectrum of local non-Gaussianity may be derived:

\begin{eqnarray}
B_\Psi (\nk_1,\nk_2,\nk_3) &=& 2 f_{\textrm{NL}}^\textrm{loc} [ P_\Psi (\nk_1) P_\Psi (\nk_2) \\ \nonumber
&+& P_\Psi (\nk_2) P_\Psi (\nk_3) + P_\Psi (\nk_3) P_\Psi (\nk_1) ].
\end{eqnarray}

In the standard picture, the non-Gaussianity produced by many single-field slow-roll models  is considered small and likely unobservable.
 However, a relatively  large,  possibly detectable, amount of non-Gaussianity can be  expected  when any of the following conditions are violated \cite{34,36,42,43}:

\begin{enumerate}

\item \emph{Single Field}. There was only one quantum field responsible for driving inflation

\item \emph{Canonical Kinetic Energy}. The kinetic energy of the quantum field is such that the speed of propagation of fluctuations is equal to the speed of light.

\item \emph{Slow Roll}. The evolution of the field was always very slow compared to the Hubble time during inflation

\item \emph{Initial Vacuum State}. The quantum field was in the preferred ``Bunch-Davies vacuum" state.

\end{enumerate}

\subsection{Non-Gaussianity in the CMB}\label{obs}

In this subsection, we present the standard connection between the primordial bispectrum at the end of inflation and the observed bispectrum of CMB anisotropies.

\subsubsection{Theoretical predictions for the CMB bispectrum from inflation}

As we mentioned in Sec. \ref{Intro}, the temperature anisotropies are represented using the $\alpha_{lm}$ coefficients of a spherical harmonic decomposition of the celestial sphere,

\beq
\frac{\delta T}{T_0} (\theta, \varphi) = \sum_{lm} \alpha_{lm} Y_{lm} (\theta, \varphi),
\eeq
and the curvature perturbation $\Psi$ is imprinted on the CMB multipoles $\alpha_{lm}$ by a convolution involving the called transfer functions $\Delta (k)$ representing the linear perturbation evolution, through Eq. \eqref{alm2}:

\begin{equation*}
\alpha_{lm} = \frac{4 \pi i^l}{3}   \int \frac{d^3{k}}{(2 \pi)^3} j_l (kR_D) Y_{lm}^* (\hat{k}) \Delta (k)
%(\eta_R,\eta_D)
 \Psi_{\vec{k}} (\eta_R).
\end{equation*}

The CMB bispectrum, also called the \emph{angular bispectrum}, is defined as the three-point correlator of the $\alpha_{lm}$:

\beq\label{angbis}
B_{m_1 m_2 m_3}^{l_1 l_2 l_3} \equiv \overline{\alpha_{l_1 m_1} \alpha_{l_2 m_2} \alpha_{l_3 m_3}}.
\eeq

At this point, the standard picture leads us to another subtle issue: that is, the overline in Eq. \eqref{angbis} denotes, in principle, an average over an ensemble of Universes. In reality, we cannot measure the ensemble average of the angular harmonic spectrum,  as  we have access to just one realization, say, the collection of complex  numbers: $\{ \alpha_{l_1 m_1}, \alpha_{l_2 m_2}, \ldots, \alpha_{l_n m_n} \}$. In order to overcome this issue, the standard approach relies on the {\it ergodic assumption} \cite{27}. The ergodicity of a system refers to that property of  a process by which the average value of a system's characteristic, measured over time, is the same as the average value measured over an appropriately constructed  ensemble. If one accepts the common supposition that the inflationary perturbation is indeed ergodic, then one expects the volume average of the fluctuations to behave like the ensemble average: The Universe may contain regions where the fluctuation is atypical, but with high 
probability most regions contain fluctuations with a root-mean-square amplitude close to $\sigma$ \cite{44}. Therefore, the probability distribution on the ensemble, which is encoded in \eqref{angbis}, translates to a probability distribution on smoothed regions of a determined size within our own Universe.

After the above analysis, we continue with the calculation relating the primordial bispectrum with the angular bispectrum. By substituting Eq. \eqref{alm2} in Eq. \eqref{angbis}, one obtains

\begin{widetext}

\begin{eqnarray}\label{fullbis}
B_{m_1 m_2 m_3}^{l_1 l_2 l_3} &=& \bigg( \frac{4\pi}{3} \bigg)^3 i^{l_1 + l_2 + l_3} \int \frac{d^3 k_1}{(2\pi)^3} \frac{d^3 k_2}{(2\pi)^3} \frac{d^3 k_3}{(2\pi)^3} \Delta(k_1) \Delta (k_2)  \Delta(k_3) j_{l_1} (k_1 R_D) j_{l_2} (k_2 R_D) j_{l_3} (k_3 R_D) \overline{\Psi_{\nk_1} \Psi_{\nk_2} \Psi_{\nk_3} } \nonumber \\ 
&\times& Y_{l_1 m_1} (\hat{k_1} ) Y_{l_2 m_2} (\hat{k_2} ) Y_{l_3 m_3} (\hat{k_3} ) \nonumber \\
&=& \bigg( \frac{2}{ 3 \pi} \bigg)^3  \int dk_1 dk_2 dk_3 \textrm{ } (k_1 k_2 k_3)^2 B_\Psi (k_1,k_2,k_3) \Delta (k_1) \Delta (k_2)  \Delta(k_3) \times j_{l_1} (k_1 R_D) j_{l_2} (k_2 R_D) j_{l_3} (k_3 R_D) \nonumber \\
&\times& \int_0^\infty dx \textrm{ } x^2 j_{l_1} (k_1 x) j_{l_2} (k_2 x) j_{l_3} (k_3 x)  \int d \Omega_{\hat{x}} Y_{l_1 m_1} (\hat{x} ) Y_{l_2 m_2} (\hat{x} ) Y_{l_3 m_3} (\hat{x} ),
\end{eqnarray}
where in the last line, we have integrated over the angular parts of the three $k_i$ and used the exponential integral form for the delta function that appears in the bispectrum definition \eqref{bispec}. The last integral over the angular part of $\x$ is known as the Gaunt integral, which can be expressed in terms of Wigner 3-$j$ symbols as

\begin{eqnarray}
\mathcal{G}_{l_1 l_2 l_3}^{m_1 m_2 m_3} &\equiv& \int d \Omega_{\hat{x}} Y_{l_1 m_1} (\hat{x} ) Y_{l_2 m_2} (\hat{x} ) Y_{l_3 m_3} (\hat{x} ) \nonumber \\
&=& \sqrt{\frac{(2l_1+1)(2l_2+1)(2l_3+1) }{ 4\pi } } \left( \begin{array}{lcr}
      l_1& l_2 & l_3  \\
     0 & 0 & 0
    \end{array}
    \right) \left( \begin{array}{lcr}
      l_1& l_2 & l_3  \\
     m_1 & m_2 & m_3
    \end{array}
    \right).
\end{eqnarray}

\end{widetext}

The fact that the bispectrum $B_{m_1 m_2 m_3}^{l_1 l_2 l_3}$ consists of the Gaunt integral, $\mathcal{G}_{l_1 l_2 l_3}^{m_1 m_2 m_3}$, implies that the bispectrum satisfies the triangle conditions and parity invariance: $m_1 + m_2 + m_3 = 0$, $l_1 + l_2 + l_3 =$ even, and $|l_i-l_j| \leq l_k \leq l_i +l_j$ for all permutations of indices.

One, thus, can write

\beq\label{reduced}
B_{m_1 m_2 m_3}^{l_1 l_2 l_3} = \mathcal{G}_{l_1 l_2 l_3}^{m_1 m_2 m_3} b_{l_1 l_2 l_3},
\eeq
where $b_{l_1 l_2 l_3}$ is an arbitrary real symmetric function of $l_1, l_2$, and $l_3$. This form, Eq. \eqref{reduced}, is necessary and sufficient to construct generic $B_{m_1 m_2 m_3}^{l_1 l_2 l_3}$ satisfying  rotational invariance; thus, in the literature, one encounters $b_{l_1 l_2 l_3}$ more frequently than $B_{m_1 m_2 m_3}^{l_1 l_2 l_3}$. The quantity $b_{l_1 l_2 l_3}$ is called the \emph{reduced} bispectrum, as it contains all the physical information in $B_{m_1 m_2 m_3}^{l_1 l_2 l_3}$. Since the reduced bispectrum does not contain the Wigner 3-$j$ symbol, which merely ensures the triangle conditions and parity invariance, it is easier to calculate  the physical properties of the theoretical bispectrum.

In the standard picture, one assumes that, if there is a nontrivial bispectrum, then it has arisen through a physical process that is statistically isotropic\footnote{Although, it would be interesting, and possibly a more realistic approach to the problem, to proceed in the analysis without this assumption.}, so we can employ the angle-averaged bispectrum $B_{l_1 l_2 l_3}$ without loss of information, that is \cite{33,34},

\beq\label{avg}
B_{l_1 l_2 l_3} = \sum_{m_i}  \left( \begin{array}{lcr}
      l_1& l_2 & l_3  \\
     m_1 & m_2 & m_3
    \end{array}
    \right) \overline{\alpha_{l_1 m_1} \alpha_{l_2 m_2} \alpha_{l_3 m_3}}.
\eeq
We now can obtain a relation between the averaged bispectrum, $B_{l_1 l_2 l_3}$ and the reduced bispectrum $b_{l_1 l_1 l_2}$, by substituting Eq. \eqref{reduced} into Eq. \eqref{avg},

\beq\label{avgred}
B_{l_1 l_2 l_3 } = \sqrt{\frac{(2l_1+1)(2l_2+1)(2l_3+1) }{ 4\pi } } \left( \begin{array}{lcr}
      l_1& l_2 & l_3  \\
     0&0 & 0
    \end{array}
    \right) b_{l_1 l_2 l_3},
\eeq
where the identity

\begin{eqnarray}
&\sum_{\textrm{all } m}& \left( \begin{array}{lcr}
      l_1& l_2 & l_3  \\
     m_1& m_2 & m_3
    \end{array}
    \right) \mathcal{G}_{l_1 l_2 l_3}^{m_1 m_2 m_3} \nonumber \\
    &=& \sqrt{\frac{(2l_1+1)(2l_2+1)(2l_3+1) }{ 4\pi } } \left( \begin{array}{lcr}
      l_1& l_2 & l_3  \\
     0&0 & 0
    \end{array}
    \right) \nonumber \\
\end{eqnarray}
was used. The reduced bispectrum obtained from Eq. \eqref{fullbis} then takes the much simpler form

\begin{eqnarray}\label{red2}
b_{l_1 l_2 l_3} &=& \bigg( \frac{2}{3 \pi} \bigg)^3  \int dk_1 dk_2 dk_3 \textrm{ } (k_1 k_2 k_3)^2 B_\Psi (k_1, k_2, k_3)  \Delta (k_1) \nonumber \\ 
&\times& \Delta (k_2)  \Delta (k_3)  j_{l_1} (k_1 R_D) j_{l_2} (k_2 R_D) j_{l_3} (k_3 R_D) \nonumber \\
&\times& \int_0^\infty dx \textrm{ } x^2 j_{l_1} (k_1 x) j_{l_2} (k_2 x) j_{l_3} (k_3 x).
\end{eqnarray}

This is the main equation for this section, since it explicitly relates the primordial bispectrum, predicted by the standard inflationary theories, to the averaged bispectrum through Eq. \eqref{avgred} obtained from the CMB angular bispectrum $\overline{\alpha_{l_1 m_1} \alpha_{l_2 m_2} \alpha_{l_3 m_3}}$. This formula is entirely analogous to the well-known relation linking the primordial power spectrum $P_\Psi(k)$ and the CMB angular power spectrum $C_l$, i.e.

\beq
C_l = \frac{2}{9 \pi} \int k^2 P_\Psi(k) \Delta^2 (k) j_l^2 (k R_D) dk.
\eeq

\subsubsection{Measuring primordial non-Gaussianity from the CMB}

As we mentioned before, in most inflationary models, the parameter characterizing primordial non-Gaussianity is $f_{\textrm{NL}}$.
 Thus, the next task within the standard picture is to estimate $f_\textrm{NL}$ from the CMB data set. That is, one chooses the primordial model that one wants to test, characterizing it through its bispectrum shape. One then proceeds to estimate the corresponding amplitude $f_\textrm{NL}^\textrm{model}$ from the data. If the final estimate is consistent with $f_\textrm{NL}^\textrm{model} = 0$, one concludes that no significant detection of the given shape is produced by the data, but one still determines important constraints on the allowed range of $f_\textrm{NL}^\textrm{model}$.
 Note that, ideally, one would like to do more than just constrain the overall amplitude and reconstruct the entire shape from the data by measuring configurations of the bispectrum. However, the expected primordial signal is too small to allow the signal from a single bispectrum triangle to emerge over the noise. For this reason, one studies the cumulative signal from all the configurations that are sensitive to $f_\textrm{NL}^\textrm{model}$.

Given the above analysis, the standard picture then makes use of estimation theory to extract an estimate for $f_\textrm{NL}$ from the CMB data set. An unbiased bispectrum-based minimum variance estimator for the nonlinearity parameter can be written as \cite{45,46}

\begin{eqnarray}
\label{est}
\hat{f}_\textrm{NL} &=& \frac{1}{N} \sum_{l_i m_i} \left( \begin{array}{lcr}
      l_1& l_2 & l_3  \\
     m_1& m_2 & m_3
    \end{array}
    \right)  \frac{B_{l_1 l_2 l_3}^\textrm{th}}{(C_{l_1} C_{l_2} C_{l_3})_\textrm{obs}} \nonumber \\
    &\times& (\alpha_{l_1 m_1} \alpha_{l_2 m_2} \alpha_{l_3 m_3})_\textrm{obs},
\end{eqnarray}
where $B_{l_1 l_2 l_3}^\textrm{th}$ is the angle-averaged theoretical CMB bispectrum for the model in consideration, with $f_{\textrm{NL}}^\textrm{th}=1$; $C_l$ is the observed angular spectrum; and $\alpha_{lm}$ are the multipoles of the observed CMB temperature fluctuations. The normalization $N$ is  calculated  requiring  the estimator to be ``unbiased,'' i.e. the averaged value is equal to the ``true" value of the parameter, $\langle \hat{f}_\textrm{NL} \rangle = f_\textrm{NL}$. If the bispectrum $B_{l_1 l_2 l_3}$ is calculated for $f_{NL} = 1$, then the normalization takes the following form

\beq
N= \sum_{l_i} \frac{ (B_{l_1 l_2 l_3})^2}{C_{l_1} C_{l_2} C_{l_3}}.
\eeq

The estimator for non-Gaussianity \eqref{est} is then  simplified using Eqs. \eqref{red2} and \eqref{avgred} to yield

\begin{eqnarray}\label{NL}
\hat{f}_\textrm{NL} &=& \frac{1}{N} \sum_{l_i m_i} \int d\Omega_{\hat{x}} Y_{l_1 m_1} (\hat{x}) Y_{l_2 m_2} (\hat{x}) Y_{l_3 m_3} (\hat{x}) \nonumber \\
&\times& \int_0^\infty x^2 dx j_{l_1} (k_1 x) j_{l_2} (k_2 x) j_{l_3} (k_3 x)  (C_{l_1}^{-1} C_{l_2}^{-1} C_{l_3}^{-1})_\textrm{obs}  \nonumber \\
&\times&  \bigg(\frac{2}{\pi} \bigg)^3\int  dk_1  dk_2  dk_3 (k_1 k_2 k_3)^2 B(k_1,k_2,k_3) \Delta(k_1) \nonumber \\
&\times& \Delta (k_2) \Delta (k_3) j_{l_1} (k_1 R_D) j_{l_2} (k_2 R_D) j_{l_3} (k_3 R_D) \nonumber \\
&\times& (\alpha_{l_1 m_1} \alpha_{l_2 m_2} \alpha_{l_3 m_3})_\textrm{obs},
\end{eqnarray}
where $B(k_1,k_2,k_3)$ is the primordial bispectrum obtained from the three-point function, as defined in Eq. \eqref{bispec}.   In this  manner, the sought  constraints  are obtained.   The best results, corresponding  to the so-called, local, equilateral, and orthogonal shape of non-Gaussianities using the WMAP 7 year data \cite{47},  yield $f_\textrm{NL}^\textrm{local}$ = 32 $\pm$ 21 (1$\sigma$), $f_\textrm{NL}^\textrm{equil}$ = 26 $\pm$ 140 (1$\sigma$), and $f_\textrm{NL}^\textrm{orthog}$ = -202 $\pm$ 104 (1$\sigma$).

\section{ The collapse  model  account for   the inflationary   origin of  cosmic structure}
\label{colapso}
\smallskip

Before proceeding, it seems worthwhile to briefly
explain the view we take regarding quantum physics
and Einstein's gravity. The framework we adopt is
based on a description of the problem that allows, at
the same time, the quantum treatment of other fields
and a classical treatment of gravitation, that is, the
realm of semiclassical gravity, together with quantum
field theory in curved space-time. We will assume that
to be a valid approximation most of the time, with the
exception associated precisely with the dynamical
collapse, as we will explain below. Such a description
of gravitation in interaction with quantum fields is
characterized by the semiclassical Einstein equation:
$R_{\mu \nu} - (1/2)g_{\mu \nu} R = 8 \pi G \bra \hat{T}_{\mu \nu} \ket$,
 whereas the other fields,
including the inflaton, are treated in the standard quantum field theory fashion. It seems clear that this approximated description would break down in association
with the quantum mechanical collapses or state jumps,
which we are considering to be part of the underlying
quantum theory containing gravitation. The reason for
this breakdown is simply that the left-hand side of the
equation above has zero divergence ($\nabla_{\mu} G^{\mu \nu} = 0$), while
the divergence of the right-hand side, $\nabla_{\mu} \bra \hat{T}^{\mu \nu} \ket$, will be
nonvanishing (even discontinuous) in connection with
the jumps of the quantum state (such a jump is how we
are describing here the self-induced collapse of the
wave function).

In this setting, we start from the assumption that, in
accordance with the standard inflationary accounts, and
as mentioned before, the state of the Universe before the
time at which the seeds of structure emerge is described by the H\&I Bunch-Davies vacuum state for the matter degrees of freedom (DOF) and the corresponding H\&I classical Robertson-Walker space-time.

Then, we assume that, at a later stage, the quantum state
of the matter fields reaches a stage whereby the corresponding state for the gravitational DOF is forbidden,
and a quantum collapse of the matter field wave function
is triggered by some unknown physical mechanism. In this
manner, the state resulting from the collapse of the quantum state of the matter fields does not need to share the
symmetries of the initial state. After the collapse, the
gravitational DOF are assumed to be, once more, accurately described by Einstein's semiclassical equation.
However, as $ \bra \hat{T}_{\mu \nu} \ket$for the new state does not need to
have the symmetries of the precollapse state, we are led
to a geometry that, generically, will no longer be homogeneous and isotropic.

The starting point of the specific analysis is the same as
the standard picture, i.e., the action of a scalar field coupled
to gravity:

\beq
\label{eq_action}
S=\int d^4x \sqrt{-g} \lbrack {\frac{1} {16\pi G}} R[g] - 1/2\nabla_a\phi
\nabla_b\phi g^{ab} - V(\phi)\rbrack,
\eeq
 where $\phi$ stands for the inflaton and $V$ stands for the
inflaton's potential.
 One then splits both metric and
scalar fields into a spatially homogeneous part (``background'') and an
inhomogeneous part (``fluctuation''), i.e. $g=g_0+\delta g$,
$\phi=\phi_0+\delta\phi$.

The background is  taken to  be the spatially flat Friedmann-Robertson Universe  with line element
$ ds^2=a(\eta)^2\left[- d\eta^2 + \delta_{ij} dx^idx^j\right]$ and the homogeneous scalar field $\phi_0(\eta)$. The  evolution equations  for this  background are
scalar field equations,
\begin{equation}
\ddot\phi_0 + 2 \frac{\dot a}{ a}\dot\phi_0 +
a^2\partial_{\phi}V[\phi] =0,
 \qquad
3\frac{\dot a^2}{a^2}=4\pi G  (\dot \phi^2_0+ 2 a^2 V[\phi_0]).
\end{equation}
The  scale factor 
 corresponding  to the inflationary regime,  written in terms of the conformal time, is:
$a(\eta)=-1/[H_I^2(1-\epsilon)\eta]$
 with $ H_I ^2\simeq  (8\pi/3) G V$. The slow-roll parameter  $\epsilon \equiv 1-\dot{\mathcal{H}}/\mathcal{H}^2$  is considered to be very small $\epsilon \ll 1$ during the inflationary stage. The Hubble factor $H_I$ is approximately constant, and the scalar field $\phi_0$ is in the slow roll regime, i.e., $\dot\phi_0= - ( a^3/3 \dot a) \partial_\phi V$.
  According to   the standard  inflationary scenario, this era is followed by  a reheating period in which the Universe is repopulated with ordinary matter fields,
   a regime that  then evolves toward a standard hot
 big bang cosmology regime leading up to the present cosmological time. The functional  form  of $a(\eta)$ during these latter periods changes, but we
  will ignore those details because most of the change in the value of $a$ occurs during the inflationary regime.
  We  will set  $ a=1$ at the ``present  cosmological time",  and assume that  the inflationary regime ends at a value of $\eta=\eta_0$, negative and very small  in absolute terms ($\eta_0 \simeq -10^{-22}$ Mpc).

   Next, we turn to consider  the perturbations. We  shall focus in this  work  on the   scalar perturbations,  and ignore,  for simplicity,
   the tensor perturbations or  gravitational waves.  Working in  the  so-called  longitudinal
    gauge, the perturbed metric is  written as:
\begin{equation}
ds^2=a(\eta)^2\left[-(1+ 2 \Psi) d\eta^2 + (1- 2
\Psi)\delta_{ij} dx^idx^j\right],
\end{equation}
 where $\Psi$  stands for the scalar perturbation usually known as
the \emph{Newtonian potential}.

The perturbation of the scalar field is  related to a perturbation of the energy-momentum tensor and
reflected  into  Einstein's equations,  which, at the lowest order,  lead to the following   constraint equation for the  Newtonian potential:
\begin{equation}
\nabla^2 \Psi  =4\pi G \dot \phi_0  \delta\dot\phi= s   \delta\dot\phi,
\label{main3}
\end{equation}
where we introduced the abbreviation $s \equiv 4\pi G \dot \phi_0$.

Now, we consider in some detail  the quantum theory of the field $\delta\phi$.
It is convenient  to  work with the  rescaled field variable  $y=a \delta \phi$ and its conjugate momentum $ \pi = \dot{y} - y\dot{a}/a$.
 For  simplicity, we set the problem in  a finite   box of side $L$,  which can be taken to $ \infty$  at the end of all calculations.  We
   decompose the
field  and momentum operators  as
\begin{equation}
\begin{split}
\hat{y}(\eta,\vec{x}) =
 \frac{1}{L^{3}}\sum_{\vec k}\ e^{i\vec{k}\cdot\vec{x}} \hat y_{\nk}
(\eta), \\
\py(\eta,\vec{x}) =
\frac{1}{L^{3}}\sum_{\vec k}\ e^{i\vec{k}\cdot\vec{x}} \hat \pi_{\nk}
(\eta),
\end{split}
\end{equation}
where the sum is over the wave vectors $\vec k$ satisfying $k_i L=
2\pi n_i$ for $i=1,2,3$, with $n_i$ integer, and where $\hat y_{\nk} (\eta) \equiv y_k(\eta) \ann_{\nk} + y_k^*(\eta)
\cre_{-\nk}$ and  $\hat \pi_{\nk} (\eta) \equiv g_k(\eta) \ann_{\nk} + g_{k}^*(\eta)
\cre_{-\nk}$
with the usual  choice of modes:
 \beq
 \begin{split}
 y_k(\eta) = \frac{1}{\sqrt{2k}}\left(1 - \frac{i}{\eta
k}\right)\exp(- i k\eta), \\
g_k(\eta) = -
i\sqrt{\frac{k}{2}}\exp(- i k\eta),
\end{split}
\eeq
 which  leads to  what is known  as the Bunch-Davies  vacuum.

 Note  that, according to the point  of  view  we  discussed at the beginning of this section  and  having, at  this point, the quantum theory for the relevant matter fields,
  the effects  of the quantum fields   on the geometrical variables  are codified  in the  semiclassical Einstein equations.  Thus  Eq. (\ref{main3})  must be replaced  by

\begin{equation}
\nabla^2 \Psi  =4\pi G \dot \phi_0  \delta\dot\phi= s \l \hat{\dot{\delta \phi}} \r = (s/a) \l \hat{\pi} \r.
\label{SemiEE}
\end{equation}
At this point,   one  can  clearly observe that,
 if  the  state of the  quantum field  is in the vacuum state,  the metric
  perturbations  vanish and thus  the space-time is homogeneous and isotropic.

As  already mentioned, our  proposal  is  based on  the consideration of  a self-induced collapse, which
  we  take  to  operate in close analogy with  a ``measurement" (but  evidently, with no external  measuring apparatus  or observer   involved). This  leads us to want  to work
 with  Hermitian operators, which in ordinary quantum mechanics are the ones susceptible to direct measurement.
 Therefore,  we must  separate  both $\hat y_{\nk} (\eta)$ and $\hat \pi_{\nk}
(\eta)$ into their real and imaginary parts $\hat y_{\nk} (\eta)=\hat y_{\nk}{}^R
(\eta) +i \hat y_{\nk}{}^I (\eta)$ and $\hat \pi_{\nk} (\eta) =\hat \pi_{\nk}{}^R
(\eta) +i \hat \pi_{\nk}{}^I (\eta)$
so that the operators $\hat y_{\nk}^{R, I} (\eta)$ and $\hat
\pi_{\nk}^{R, I} (\eta)$ are
 Hermitian operators.

So  far,  we have  proceeded  in  a  manner  similar  to the standard  one,   except in that we are  treating at the quantum level  only the scalar field and  not the  metric  fluctuation.    At this point it is   worthwhile  to   emphasize  that the  vacuum state   defined  by $ \ann_{\nk}{}
^{R,I} |0\r =0$  is  100\%  translational  and rotationally invariant (see Appendix \ref{BD}).

 For the  next step, we must   specify  in more detail  the   modeling of  the  collapse.  Then, we  must take   into account that, after   the collapse has  taken place,  one should   consider the  continuing  evolution  of the  expectation values of the  field  variables
 until  the end of inflation, and eventually  up to the hypersurface of decoupling.  In fact,  if we  wanted to actually   compare  our analysis  directly with  observations,  we would need  to evolve  the  perturbations both through  the  reheating  period  and  through the decoupling era. This,  however,  is   normally  taken into account  through the use of appropriate transfer functions, and  we  will assume that  the  same  procedure could   be implemented  in the context of   the present analysis, but we  will not consider it  further in the  present manuscript.

We will further assume that the collapse is somehow analogous to an imprecise measurement\footnote{An imprecise measurement of an observable is one in which one does not end up with an exact eigenstate  of that observable but  rather with a state that is  only peaked around the eigenvalue. Thus, we could consider measuring a  certain particle's position and momentum so as to end up with a state that is a wave packet with both position and momentum defined to a limited extent and which,  of course, does not  entail a conflict with Heisenberg's uncertainty bound.}
of the
operators $\hat y_{\nk}
^{R, I}
(\eta)$ and $\hat \pi_{\nk}
^{R, I}
(\eta)$.
Now, we will specify the rules according to which collapse happens.
Again, at this point our criteria will be simplicity and naturalness.
What we have to describe is the state $|\Theta\rangle$ after the
collapse. 

% It turns out that, for the goals at hand, all we need to specify  is the  quantity
%$d^{R,I}_{\nk} \equiv \langle\Theta|\ann_{\nk}^{R,I}
%|\Theta\rangle $, as this  determines   the  expectation value of the field and momentum  operator for the mode  $\nk$ at all times after the collapse.

 It seems  natural to assume (taking the view that a collapse  effect on  a state is analogous  to some  sort of  approximate measurement) that  after the  collapse, the expectation values of the field and momentum operators  in each mode  will  be related to the uncertainties  of the  precollapse  state (recall that the  expectation  values  in the vacuum state  are  zero). In the vacuum state, $\hat{y}_{\nk}$ and
$\hat{\pi}_{\nk}$ individually are distributed according to Gaussian
wave functions centered at 0 with spread $\fluc{\hat{y}_{\nk}}_0$ and
$\fluc{\hat{\pi}_{\nk}}_0$, respectively.

  We  might   consider  various possibilities   for the detailed  form  of this collapse. Thus, for   their generic form,
  associated with the  ideas  above, we write

\begin{equation}
\l {\hat{y}_{\nk}^{R,I}
(\eta^c_k)} \r_\Theta= \lambda_1 x^{R,I}_{\nk,1}
\sqrt{\fluc{\hat{y}^{R,I}_{\nk}}_0}
=\lambda_1 x^{R,I}_{\nk,1}|y_k(\eta^c_k)|\sqrt{\hbar L^3/2},
\label{momentito}
\end{equation}

\begin{equation}
\l {\hat{\pi}_{\nk}{}^{R,I}
(\eta^c_k)}\r_\Theta
= \lambda_2 x^{R,I}_{\nk,2}\sqrt{\fluc{\pyRI_{\nk}}
_0} ,
=\lambda_2 x^{R,I}_{\nk,2}|g_k(\eta^c_k)|\sqrt{\hbar L^3/2},
\label{momentito1}
\end{equation}
where $x_{\nk,1}^{R,I}
,x_{\nk,2}^{R,I}
$ have been  assumed, in our previous  works, to be   selected randomly from within a Gaussian
distribution centered at zero with spread one, and $\tc$ represents the \emph{time of collapse} for each mode. Here,   $\lambda_1$ and $\lambda_2$  are  parameters taking the  values   0 or 1 that allow us to specify the  kind of  collapse proposal we want to consider (The  main ones  we  have considered  in \cite{4,8,9} are   $\lambda_1= \lambda_2 =1  $  for the {\it symmetric  collapse}  and $\lambda_1= 0,  \lambda_2 =1  $  for the {\it Newtonian collapse}). At this point, we must emphasize that our Universe corresponds
to a single realization of these random variables, and, thus,  each of these quantities
$ x^{R,I}_{\nk,1}$, $ x^{R,I}_{\nk,2}$ has a  single specific value.    The  fact that we  can  represent the  specific  details of the  first inhomogeneities  and anisotropies, the  seeds of cosmic  structure, is something that has no  counterpart on the  standard treatments.  It is  clear that  one can  now investigate  how  the different  specific  proposals  for the   process of collapse  could   affect the statistics  of the  $x_{\nk,1}^{R,I}
,x_{\nk,2}^{R,I}$ . One  could now   inquire  about  both, the statistics  of  these  quantities in some imaginary ensemble of possible universes as  well as the statistics of  such  quantities for the particular  Universe  we inhabit.

 Regarding the collapse models, it should be clear that  there  are  many other  possibilities that  we  have not  even thought about  and that might require  drastically  modified  formalisms.
In fact  in a recent work \cite{13} grounds were found that  suggest  a correlation between the  $ x^{R,I}_{\nk,1}$, $ x^{R,I}_{\nk,2}$  of any mode  with those of their higher harmonics (something reminiscent of the  so-called  parametric   resonances  found  in  quantum optics in materials  with  nonlinear   response  functions \cite{48}). As  we noted  in Ref. \cite{14} those  particular types  of correlations,  in  turn,   would lead  to   a  very specific  signature, which  might be  looked for in the statistical  features of the CMB.

%On the other hand, we have  been using the values for $\lambda_1$ and $\lambda_2$ to characterize the different collapse schemes, e.g.: i) $\lambda_1=0$, $\lambda_2=1$, ii) $\lambda_1=\lambda_2=1$ (which we call the ``symmetric scheme").
%It is clear that one can devise  many other models  of  collapse,  a good fraction of them  can be  described within in the  scheme above, while others require  a  slightly  modified treatment \cite{Adolfo}.  Still,   there are surely   many other  possibilities which  we  have not  even thought about  and which might require  drastically  modified  formalisms.

Returning to the specific  models we have  described  above,   we need to compute the relevant  expectation  values of the field operators in the post-collapse state $|\Theta\rangle$ at the relevant times.
 For  each  specific   model we  do this  by
using   Eqs. \ref{momentito}  and  \ref{momentito1} above
   %one can solve in each case  for  the quantities
  % $d^{R,I} _{\nk}$,
 and   the evolution equations  for the  expectation values (i.e.  using Ehrenfest's theorem). Thus one  obtains
$\l {\hat{y}_{\nk}{}^{R,I} (\eta)} \r$ and
 $\l {\hat{\pi}_{\nk}{}^{R,I} (\eta)} \r $ for the   state  that resulted from the collapse, for all later times. The explicit expressions for the $\bra \hat{y}_{\nk}^{R,I} (\eta) \ket_\Theta$ and $\bra \hat{\pi}_{\nk}^{R,I} (\eta) \ket_\Theta$ are

\bea\label{expecyeta}
\bra \hat{y}_{\nk}^{R,I} (\eta) \ket_\Theta &=& \bigg[ \frac{\cos D_k}{k} \bigg( \frac{1}{k\eta} - \frac{1}{z_k} \bigg) + \frac{\sin D_k}{k} \bigg(\frac{1}{k\eta z_k} + 1 \bigg) \bigg]  \nonumber \\ 
&\times& \bra \hat{\pi}_{\nk}^{R,I} (\tc) \ket_\Theta + \bigg( \cos D_k - \frac{\sin D_k}{k\eta} \bigg)  \bra \hat{y}_{\nk}^{R,I} (\tc) \ket_\Theta, \nonumber \\
\eea

\bea\label{expecpieta}
\bra \hat{\pi}_{\nk}^{R,I} (\eta) \ket_\Theta  &=& \bigg( \cos D_k +\frac{\sin D_k}{z_k} \bigg) \bra \hat{\pi}_{\nk}^{R,I} (\tc) \ket_\Theta \nonumber \\
&-& k \sin D_k  \bra \hat{y}_{\nk}^{R,I} (\tc) \ket_\Theta,
\eea
where $D_k \equiv k\eta - z_k$ and $z_k \equiv k\tc$. This calculation is explicitly done in Refs. \cite{4,11}

   With  this information at  hand,  we can now  compute the perturbations of the metric  after the collapse of  all the modes\footnote{In fact,  we need only  be concerned with the relevant modes, those  that  affect the observational quantities in a relevant  way. Modes that  have  wavelengths that  are either too large or too  small are irrelevant in this  sense.}.

\subsection{Connection  to  Observations}

Now, we must put together our semiclassical  description  of the gravitational
 DOF and the quantum mechanics description of the inflaton field.  We recall that this entails the semiclassical version of
the perturbed Einstein's equation that, in our case, leads to Eq.
(\ref{SemiEE}).
The Fourier components  at  the conformal time $ \eta$ are given by
\begin{equation}
 \Psi_{\nk} ( \eta) = -\sqrt{\frac{\epsilon}{2}} \frac{H_I \hbar}{M_P k^2} \l {\hat{\pi}_{\nk}{}(\eta)}  \r,
\label{SemiEEK}
\end{equation}
where we  have used the fact that,  during inflation, $ s  = \sqrt{\epsilon/2}(a H_I / M_P)$, with $M_P$ as the reduced Planck mass $M_P^2 \equiv \hbar^2/(8 \pi G)$.
The expectation value depends  on  the  state of the quantum field; therefore,   as  we  already noted, prior to the collapse, we have  $\Psi_{\nk} ( \eta)  =0$, and the space-time is  still homogeneous and isotropic at the corresponding scale. However, after the collapse  takes place,  the state of the field  is a different state
with new expectation values that  generically  will not vanish, indicating that, after this time, the Universe becomes anisotropic and inhomogeneous at the corresponding scale.
 We now can reconstruct the space-time value of the Newtonian potential using
\begin{equation}
\Psi(\eta,\vec{x})=
 \frac{1}{L^{3}}\sum_{\vec k}\ e^{i\vec{k}\cdot\vec{x}} \Psi_{\nk}
(\eta),
\label{Psi}
 \end{equation}
to extract the quantities of observational interest.

 In order to connect with the observations, we shall relate the expression \eqref{SemiEEK} for the evolution of
the Newtonian potential during the early phase of accelerated expansion to
the small anisotropies observed in the temperature of the cosmic microwave background radiation,
 $\delta T(\theta,\varphi)/T_0$ with $T_0\approx 2.725 K$ as the temperature average. %on the celestial sphere
They are considered the fingerprints of the small perturbations pervading the Universe at the time of decoupling, and undoubtedly any
model for the origin of the seeds of cosmic structure should account for them.
As already mentioned in Sec. \ref{Intro}, these  data can  be  described  in terms of the coefficients  $\alpha_{lm}$
of the multipolar series expansion, i.e., Eq. \eqref{expansion.alpha}.
The different multipole numbers $l$ correspond to different angular scales: low $l$ to large scales and high $l$ to small scales.
At large angular scales ($l \lesssim 20$) the Sachs-Wolfe effect is the dominant source for the anisotropies in the CMB.
That effect relates the anisotropies in the temperature observed today on the celestial sphere to the inhomogeneities in the Newtonian potential on the last scattering surface,
\begin{equation}\label{deltaT2}
\frac{\delta T}{T_0} (\theta,\varphi) = \frac{1}{3} \Psi (\eta_D, \vec{x}_D).
\end{equation}
Here, $\eta_D$ is the conformal
time of decoupling that lies in the matter-dominated epoch, and $\vec{x}_D = R_D (\sin \theta \sin \varphi, \sin \theta \cos \varphi, \cos \theta)$, with $R_D$ as the radius of the
last scattering surface. Furthermore, using Eq. \eqref{Psi} and  $e^{i \vec{k} \cdot \vec{x}_D} = 4 \pi \sum_{lm} i^l j_l (kR_D) Y_{lm} (\theta, \varphi) Y_{lm}^* (\hat{k})$, the expression (\ref{expansion.alpha}) for $\alpha_{lm}$ can be rewritten in the form \eqref{alm2}. The transfer function $\Delta(k)$ represents the evolution of the Newtonian potential from the end of inflation $\eta_R$  to the time of decoupling $\eta_D$, i.e.  $\Psi_{\vec{k}}(\eta_D)=\Delta (k) \Psi_{\vec{k}} (\eta_R)$.

Substituting Eq. \eqref{expecpieta} in Eq. \eqref{SemiEEK} and using Eq. \eqref{momentito},  Eq. \eqref{momentito1} gives

\bea\label{psirandom}
\Psi_{\nk} (&\eta_R&) = \frac{- (L\hbar)^{3/2} \sqrt{ \epsilon} H_I }{2 \sqrt{2} M_P  k^{3/2}} \bigg[ \lambda_2 \bigg( \cos D_k + \frac{\sin D_k}{z_k} \bigg) \nonumber \\
&\times& (x_{\nk,2}^R + i x_{\nk,2}^I) - \lambda_1 \sin D_k \bigg( 1 + \frac{1}{z_k^2} \bigg)^{1/2}  (x_{\nk,1}^R + i x_{\nk,1}^I) \bigg]. \nonumber \\
\eea

Finally, using Eq. \eqref{psirandom} in Eq. \eqref{alm2} yields

\bea\label{almcol}
\alpha_{lm} &=&- \frac{ \pi i^l \hbar^{3/2} \sqrt{ 2 \epsilon} H_I }{3(Lk)^{3/2} M_P}
\sum_{\nk} \Delta(k) j_l (k R_D) Y^*_{lm} (\hat{k}) \nonumber \\
&\times& \bigg[ \lambda_2 \bigg( \cos D_k + \frac{\sin D_k}{z_k} \bigg) (x_{\nk,2}^R + i x_{\nk,2}^I) \nonumber \\
&-& \lambda_1 \sin D_k \bigg(1+ \frac{1}{z_k^2} \bigg)^{1/2} (x_{\nk,1}^R + i x_{\nk,1}^I) \bigg],
\eea
note that in Eqs. \eqref{psirandom} and \eqref{almcol}, $D_k$ is evaluated at $\eta_R$, i.e. $D_k (\eta_R) = k\eta_R - z_k$.

It is worthwhile to mention that the relation of $\alpha_{lm}$ with the Newtonian potential, as obtained in Eq. \eqref{almcol} within the collapse framework
has  no analogue in the  usual  treatments of the subject.  It  provides us  with a  clear  identification   of the  aspects  of the  analysis  where  the ``randomness"  is  located.  In this case,  it resides  in the
randomly   selected values $ x_{\nk,1}^{\textrm{R,I}}$, $ x_{\nk,2}^{\textrm{R,I}}$  that appear in the   expressions of the collapses associated
 with  each  of the modes. Here, we also find a clarification of  how, in spite  of  the intrinsic  randomness,    we can  make  any prediction at all. The
 individual  complex quantities  $\alpha_{lm} $  correspond to  large sums of complex  contributions,  each  one  having a  certain randomness,   but leading  in
 combination [i.e. the  sum of contributions  appearing in Eq. \eqref{almcol}],  to a  characteristic  value  in  just  the same  way as a random walk made  of multiple steps. In other words,
 the justification for the use of statistics in our approach is that the quantity $\alpha_{lm}$ is the sum
of contributions from the collection of modes, each contribution being a random number leading to what is in effect,
a sort of ``two-dimensional random walk," for which the total displacement corresponds to the observational quantity.
Nothing  like  this can be found in the most popular accounts, in which the issues  we have been focussing on are
 hidden in a maze  of often  unspecified  assumptions and unjustified  identifications \cite{5}.

Thus, according to Eq. (\ref{almcol}), all the modes contribute to $\alpha_{lm}$,
with a  complex number. If we had  the outcomes   characterizing  each of the individual collapses,  we  would be  able
to  predict  the  exact   value of  each of  these individual  quantities. However, we have, at this point, no other  access to such information  than the observational quantities  $  \alpha_{lm}$ themselves.

We  hope  to be able to say something about these, but  doing so   requires the consideration  of further
  hypothesis  regarding the statistical  aspects  of  the  physics   behind the collapse as well as the conditions previous to them.

 As is generally the case  with  random walks, one cannot hope to  estimate the direction of the  final displacement.
 However, one might  say something about its estimated  magnitude.  It is for that reason  that we  will be  focusing on estimating the  most likely   value of  the  magnitude:

   \bea
|\alpha_{lm}|^2 &=& \frac{16 \pi^2}{9L^{6}} \sum_{\vec{k}, \vec{k'}} \Delta(k) \Delta(k') j_l (kR_D) j_l (k'R_D) Y_{lm}^* (\hat{k}) Y_{lm} (\hat{k'}) \nonumber \\
&\times& \Psi_{\vec{k}} (\eta_R){\Psi_{\vec{k'}}}^* (\eta_R).
\eea

 Note, however, that although in our approach, each of the quantities  $\Psi_{\vec{k}} (\eta_R)$  has, in principle, a  particular numerical value, the fact that  such  value
  is the result of a  quantum collapse characterized by random numbers  indicates we  cannot make a definite prediction for it.
 We  believe that our approach has,  among others, the  advantage of  offering  a clear  way to  express the prediction  for   the observable  quantities, in a manner in which the  aspects that are controlled  by randomness    are clearly identified. This  allows, in principle,  the  consideration  in a separate  way  of each of the hypotheses  and identifications  one  is interested in making. Our  inability of predicting the specific  values for the quantities $ |\alpha_{lm} |$,  characterizing  our observations, is then clearly identified  and located in the particular random variables  introduced in the collapse hypothesis.
 But, of course,  we  want to make  predictions. So  further considerations  become necessary, but the point is that these  are clearly identifiable. We will see  below what these hypotheses are and  how  they lead to  more specific  predictions.  One of the  advantages  we  have  is that  one is  able, in principle, to consider   removing or modifying each  one of those hypotheses.
 In this  case, we can make progress,  for instance,  by  making the  assumption that we can  regard the  specific  outcomes characterizing our Universe as  a typical member  of some  hypothetical  ensemble of universes.

     For example, we are interested in  estimating  the  most likely  value of  the magnitude of $|
   \alpha_{lm}|^2$ above,  and,   in  such  hypothetical  ensemble,  we might hope that  it comes very close to our single  sample.
    It is  worth emphasizing  that, for each $l$ and $m$,  we have one single   complex number characterizing the actual observations (and, thus, the  real  Universe  we inhabit).   For a given $l$, for instance, we should  avoid  confusing  ensemble averages   with averages of   such  quantities  over the $2l+1$ values of $m$.   The  other  Universes in the ensemble  are just  figments  of our imagination, and there is nothing in our theories that  would  indicate that they are real.

    We can  simplify things  even further  by  taking   the ensemble average $  \overline{| \alpha_{lm}|^2}$ (the bar indicates  that we are taking the  ensemble
   average)   and  identifying it with the most likely value  of the quantity, and  it is  needless to say that  these  two notions are  not exactly equal  for  many types of ensembles.
   However, let us, for the moment,  ignore this issue  and   assume the   identity of  those   two  values  and look at  the ensemble average  of the quantity $| \alpha_{lm}|^2$,   which is  given by

   \begin{widetext}

\begin{equation}
\overline{|\alpha_{lm} |^2}= \frac{16 \pi^2}{9L^{6}}
\overline{
\sum_{\vec{k}, \vec{k'}} \Delta(k) \Delta(k') j_l (kR_D) j_l (k'R_D) Y_{lm}^* (\hat{k}) Y_{lm} (\hat{k'})\Psi_{\vec{k}} (\eta_R){\Psi_{\vec{k'}}}^* (\eta_R)
}.
\end{equation}

 One can,  for instance,  assume   that   the  collapsing   events  are  all uncorrelated,  and  then consider estimating the  most likely value; thus,

\begin{equation}
\label{alm8}
| \alpha_{lm}|^2_{\textrm{ML}}= \frac{16 \pi^2}{9L^{6}}
\sum_{\vec{k}, \vec{k'}} \Delta(k) \Delta(k') j_l (kR_D) j_l (k'R_D) Y_{lm}^* (\hat{k}) Y_{lm} (\hat{k'})\overline{\Psi_{\vec{k}} (\eta_R){\Psi_{\vec{k'}}}^* (\eta_R)
}.
\end{equation}

\end{widetext}

Under the assumption
of the validity of such an approximation   and  the   additional assumption that the  random
variables $x_{\nk,1}^R, x_{\nk,1}^I, x_{\nk,2}^R, x_{\nk,2}^I$ are  all uncorrelated,  we obtain that all the information regarding the ``self-collapsing" model will be codified in the quantity

\begin{equation}\label{cantidad.importante.gausiana}
 \overline{\Psi_{\vec{k}}(\eta_R)\Psi_{\vec{k}'}^*(\eta_R)}.
\end{equation}
 Generally,  one  expects this  term to be   proportional to $ \delta_{\vec k \vec k'} $,  but  alternatives cannot be ruled  out. In fact,
  a case  in which this  assumption  is relaxed   was  explored in Ref. \cite{14}.
Furthermore,  we   will  take the  limit $-k\eta_R \to 0$ in Eq. \eqref{cantidad.importante.gausiana}, which  can be expected to be
 appropriate when  restricting interest  to  the modes that are ``outside the horizon" at the end of inflation (including the modes that give a major contribution to the  observationally relevant  quantities).

Then, with the help of Eq. \eqref{psirandom}  and after taking the continuum limit ($L \to \infty$).   we obtain
\begin{equation}
|\alpha_{lm}|^2_\textrm{ML}
=\frac{\hbar^3 \epsilon H_I^2}{36 \pi M_P^2} \int \frac {dk}{k} \Delta^2(k) j^2_l(k R_D) C(k), \label{alm4}
\end{equation}
where some of the information regarding that a collapse has occurred is contained in
 the function $C(k)$\footnote{The standard amplitude for the spectrum is usually presented as $\propto V/(\epsilon M_P^4) \propto H_I^2 /(M_P^2 \epsilon)$. The fact that $\epsilon$ is in the denominator leads, in the standard picture, to a constraint scale for $V$. However, in Eq. \eqref{alm4} $\epsilon$ is in the numerator. This is because
we have not used (and in fact we will not) explicitly the transfer function $\Delta(k)$. In the standard literature, it is common to find the power spectrum for the quantity $\zeta(x)$, a field representing the curvature perturbation in the comoving gauge. This quantity is constant for modes ``outside the horizon" (irrespectively of the cosmological epoch); thus, it avoids the use of the transfer function. The quantity $\zeta$ can be defined in terms of the Newtonian potential as $\zeta
 \equiv \Psi + (2/3)(\mathcal{H}^{-1} \dot{\Psi} + \Psi)/(1+\omega)$, with $\omega \equiv p/\rho$. For large-scale modes $\zeta_{\vec{k}} \simeq \Psi_{\vec{k}} [ (2/3) (1+\omega)^{-1} + 1]$, and during inflation $1+\omega = (2/3)\epsilon$. For these modes, $\zeta_{\vec{k}} \simeq \Psi_{\vec{k}} /\epsilon$ and the power spectrum is $\mathcal{P}_{\zeta}(k) = \mathcal{P}_{\Psi}(k) / \epsilon^2 \propto H_I^2/(M_P^2 \epsilon)  \propto V/(\epsilon M_P^4)$, which contains the correct amplitude. For a detailed discussion regarding the amplitude within the collapse framework, see Ref. \cite{10}.}. The explicit form of $C(k)$  for the class of  collapse  schemes  considered here  is
\beq
C(k)=\lambda_1^2 \bigg( 1 + \frac{1}{z_k^2} \bigg) \sin^2 z_k + \lambda_2^2 \bigg( \cos z_k - \frac{ \sin z_k}{z_k} \bigg)^2.
\eeq
As  we  have  noted  in previous works, this   quantity  becomes  a  simple  constant if  the collapse time  happens to follow  a particular pattern  in which  the time  of collapse of the  mode $\vec k$  is  given  by $\eta^c_k =  Z/k$  with $Z$ as a constant. In fact,  the standard answer
would correspond to $C(k)=$ constant (which can be thought as an equivalent ``nearly scale-invariant power spectrum"). Thus, the result obtained for the relation between the time of collapse and the mode's frequency, i.e. $\tc k$= constant is a rather strong conclusion that could  represent   relevant information about whatever the mechanism of collapse is.
%A preliminary  study of the  effects   of small  deviations from such pattern for the ``symmetric scheme" ($\lambda_1=\lambda_2=1$);    for  the ``Newtonian  scheme" ($\lambda_1 = 0$, $\lambda_2=1$),  and  for a  third  scheme  that falls outside  the category    contemplated here,   have  been carried out  in \cite{Adolfo}.

It is  quite clear  that   if  the  time of collapse of each mode does not  adjust  exactly to the  pattern $ k \eta^c_k =  Z$,  then    the  collapse    schemes under  consideration (characterized  by the values of
 $\lambda_1,\lambda_2$),  or some  other  one  resulting  in a nontrivial function function $C(k)$,  would lead to different predictions for  the  exact form of the  spectrum, and comparing these predictions with the observations can help us to discriminate between the distinct collapse schemes. An analyses of these  issues   have   been presented   in Refs. \cite{8,9}.

We end this section by noting that the treatment of the statistical aspects in the collapse proposal is
quite different from the standard inflationary paradigm. We will deepen this discussion in the next section. However, at this point, the differences should be evident. In the standard accounts, one is going from
quantum correlation functions to classical $n$-point functions averaged over an ensemble of universes; then, one  goes to $n$-point correlation functions averaged over different regions of our own Universe,
and, finally, one relates  this last quantity with the observable $|\alpha_{lm}|^2$. These series of steps are not at all direct and they involve a lot of subtle issues that the standard picture does not provide in a
transparent way. On the other hand, within  the collapse approach to the subject, the observable $|\alpha_{lm}|^2$ is related to the random variables, $x_k$, through a  two-dimensional  (i.e. the result is   the sum
of  individual complex  numbers) random walk. As we mentioned, the value of $|\alpha_{lm}|^2$ corresponds then to the ``length" of the random walk. This random walk is associated to a particular realization
of a physical quantum process (i.e. the collapse of the inflaton's wave function), and  as  we  have only access to one realization--the random walk corresponding to our own Universe--the most natural
assumption (but certainly not the only one) is that the average value of the length of the possible random walks, which corresponds to $\overline{|\alpha_{lm}|^2}$, is equal  to the most likely value, i.e. to $|
\alpha_{lm}|_{\textrm{ML}}^2$, and this, in turn, is associated with the $|\alpha_{lm}|^2$ of our observable Universe.

\section{ Further Statistical  Aspects}
\label{further}

The first thing we should now note is that there are
several statistical issues at play and that, within our approach, various novel ones emerge. One central aspect is
the exact nature of the state previous to all collapses, i.e.,
the state characterizing the first stages of the inflationary
regime, and normally taken to be the Bunch-Davies vacuum. There are various possibilities that might modify the
nature of that state: For instance, if the field is not truly a
free field, and self-interactions are important, one might
find correlations between the various modes of the field.
These effects could be manifest, for instance, by nonvanishing values of quantities like $\bra 0 | \hat{y}_{\nk} \hat{y}_{\nk'} | 0 \ket  $ (as argued in the
case studied in Ref. \cite{14}). However, we should be aware
not only of the inherent problems of accessing those statistical signatures associated with the fact that we have at
our disposal a single Universe but also that our Universe,
including the relevant perturbations, is not characterized by
the vacuum state but rather by the state that results after the
collapses of all the modes, and it is quite clear that the
collapse process itself can be a source of unexpected
correlations. These would manifest themselves, for example, in correlations between the values taken by the
$x_{\nk}$'s appearing in the collapse process and which we have
so far assumed were different and independent quantities
for each mode.

Moreover, we have to note that the quantities that are
more or less directly accessible to observational investigation are not the $\bra \Theta | \hat{y}_{\nk} | \Theta \ket$, and the $n$-point functions, in
general, for the post-collapse state, but the various $\alpha_{lm}$s,
and the latter are related to the former, as can be seen in
Eq. (48) in a nontrivial way. In fact, as we saw, each $\alpha_{lm}$
corresponds to a sort of two-dimensional random walk
(i.e., a sum of complex quantities), and each of the steps
is $\bra \Theta | \hat{y}_{\nk} | \Theta \ket$. It is, thus, clear that there might be
correlations between the various $\alpha_{lm}$s simply due to the
fact that they arise from different combinations of the same
random variables. Of course, we should note that the
version of the collapse proposal we have presented here
is based on the assumption that the elementary collapse processes were associated with the observables $\hat{y}_{\nk}$
 and their
conjugate momenta according to Eqs. (40) and (41). It is
clearly conceivable that the elementary process might have
been associated, instead, with other observables. One sim-
ple possibility for those alternative observables is the various options offered by linear combinations of the former.

\subsection{The new  outlook on   non-Gaussianities}

In this section, we discuss the aspects that need modification in the study of primordial non-Gaussianities, in view
of the approach we have been discussing to the origin of
the primordial fluctuations.

The first point we should stress is that, from the two
aspects of cosmology mentioned in Sec. \ref{Intro}, we have seen
that we have had to modify the first, namely, the nature of
the quantum state, in order to be compatible with the
existence, at the fundamental (quantum) level, of the inhomogeneities and anisotropies that are behind the emergence of structure and, thus, of everything--including observers--in our Universe.

In other words, the standard physics of the very early
Universe had to be supplemented with the collapse hypothesis in order to fully account for the process that
created the primordial seeds for large-scale structure.
Otherwise, we could not really identify the process by
which the inhomogeneity and anisotropies emerged from
the initial vacuum.

As in the standard approach, we take the curvature
perturbations $\Psi$ to be the generators of the CMB anisotropy, $\delta T/T$. However, in our approach, the observed fluctuations are determined, not just by the initial vacuum
state, which is and remains homogeneous and isotropic,
but also by the characteristics of the collapse process,
besides, of course, by the effects of the late-time physics.

In this more precise and detailed approach, it is clear
that, even if the primordial state can be considered as
Gaussian, in the sense that the corresponding $n$-point
functions are completely determined by the two-point
functions--and, thus, the odd n-point functions vanish--
it might still be possible for the collapse processes to
drastically affect and modify this. In other words, there
exists, in principle, the possibility that the collapse process
itself introduces non-Gaussian characteristics into the
state. We will not discuss this possibility here, but only
point it out as something to have in mind and a topic for
future research.

As we have argued, the quantity of observational interest
is not really $\bra 0 | \hat{\Psi} (\x, \eta) \hat{\Psi} (\vec{y}, \eta) | 0 \ket$, as the argument to
justify that in the standard approach depends not only
on accepting the identification $\bra 0 | \hat{\Psi} (\x, \eta) \hat{\Psi} (\vec{y}, \eta) | 0 \ket = \overline{\Psi (\x, \eta) \Psi (\vec{y}, \eta)}$, where $\Psi(\x, \eta)$ is taken to be a classical
stochastic field and the overline denotes the average over
an ensemble of universes, but also on a series of arguments
indicating one can replace the ensemble averages with
suitable spatial averages of quantities in our Universe.

As a matter of fact, a clear example of how a careless approach to the statistics at hand can lead to
wrong conclusions is brought by the variance $\overline{\Psi^2 }$. We mentioned in Sec. \ref{stand2} that $\overline{\Psi^2 }$ diverges generically if
one does not introduce an \emph{ad hoc} cutoff for $k$. Therefore, if we consider the
temperature fluctuations in a particular point $\x_0$ of the last scattering  surface, and we estimate
it in terms of $\bra 0 | \hat{\Psi}^2 (\x_0,\eta) | 0 \ket$, we obtain a divergent quantity. Note  that we are not saying that the
temperature anisotropy is divergent, but  only that $\langle 0 | \hat{\Psi}^2 (x) |0 \rangle$, during the inflationary period, is
divergent (see Appendix \ref{div}). This  divergence  at  an  early state  would invalidate  any subsequent analysis  based on
 perturbation theory, which works  under the assumption that the  metric perturbations are small in  every point.
However,  we know from the observational data that these fluctuations of the mean temperature, in any particular
point, are rather small $\sim 10^{-5} K$. On the other hand, in the collapse proposal,  these issues  become
much less problematic  because the  scheme  indicates which variable  we  should focus on:  the variables
 subjected to the collapse are not  $\hat{y} (\x,\eta)$, $\hat{\pi} (\x,\eta)$ but the field modes $\hat{y}_{\nk} (\eta)$, $\hat{\pi}_{\nk} (\eta)$, i.e. the
 collapse does not occur in the position space, and  an independent  collapse is  assumed for each mode $\nk$. The quantities of observational
 interest, namely the $|\alpha_{lm}|^2$s, depend on the expectation values  $\bra \hat{y}_{\nk} (\eta) \ket_\Theta$, $\bra \hat{\pi}_{\nk} (\eta) \ket_\Theta$,
 in the state $|\Theta \ket$ after the collapse, and, as we  have  shown, these can be  estimated  directly  in terms of   the values of the random  variables.

As we saw   in the  introduction, if   we really took $\Psi(\x, \eta)$ to be  Gaussian and  allowed the identification of its  $n$-point functions  with the observations, we
would have to accept that  such identification holds, in  particular,  for the one-point function,  and  that  would  lead  us   to  a clear conflict between  theory and observation.

Similarly,   one  must be careful  when  we consider the    three-point function
\beq
\bra 0 | \hat { \Psi} (\x, \eta) \hat {  \Psi } (\y, \eta) \hat { \Psi} (\vec{z}, \eta) | 0 \ket,
\eeq
 with the average over an ensemble of universes $\overline{ \Psi (\x, \eta) \Psi (\y, \eta)  \Psi (\vec{z}, \eta) }$,
 and finally, the identification  of  the latter with the measured quantities as an indicator of non-Gaussian features in the cosmological perturbations.

As we saw, the   bispectrum
$\overline{ \Psi_{\nk_1} \Psi_{\nk_2} \Psi_{\nk_3}} = (2\pi)^3 \delta (\nk_1 + \nk_2 + \nk_3 )B_\Psi (k_1, k_2,k_3)$
is usually said to represent the lowest-order statistics able to distinguish non-Gaussian from Gaussian perturbations  because  Gaussianity  is  identified with
 the  requirement that  all statistical information is contained in the two-point functions and,   thus,  implicitly,  with the  vanishing  of all $n$-point functions with $n$  odd.  However,  the  lowest  odd  integer  is   not $3$ but $1$, and,  as  we  have already seen,  there is a  serious   issue that arises  when considering  the  one-point function. This,  we believe,  forces  us to question and reconsider  some  the standard arguments.

 In fact, looking anew  at the  quantities  normally associated   with the one-point function,   we  see that  we have  at our disposal, not only  the  average   quantities  $C_l$
 but  also, for every value of $l$ and $m$,  the  individual  quantities  $\alpha_{lm}$. Each one of those  corresponds, in our approach,  to different  random  walks. It  could  prove  very  interesting to   study the  distribution  of the pair of real quantities that constitute the complex number $\alpha_{lm}$:  namely, we  can
 look  at the plot  of,  say, the  real and imaginary part of  $\alpha_{lm}$, i.e. $\Re (\alpha_{lm})$ and $\Im (\alpha_{lm})$,   for a given value of $l$. This set of  $2l+1$ numbers for each one of the real and imaginary parts,  can  naturally   be  expected to display a   Gaussian   shape (which, in turn, would make the distribution of $|\alpha_{lm}|$ a Rayleigh distribution).  %Does it?

This seems to be a particularly relevant analysis, and we  do not know  of  anything like that which has  been
studied  in the literature. It  seems to us  that   the traditional approach does not naturally  lead  to  the consideration of that issue.
Looking at the   distribution of the corresponding phases  should  be
  equally  enlightening. Moreover, as   we   mentioned    in the discussion  around
  the  Eq. \eqref{almR Av}, it  would be  interesting  to  evaluate the  quantity
$\overline \alpha_{l}$  defined there, and compare the result with any  of  the
 natural estimates  for its  value,  particularly  the  expected  ensemble average of its  magnitude.

Another point worth  revisiting is that  it is  usually believed  that   a large detectable amount of non-Gaussianity can   be  expected  when the  \emph{initial  state} of the quantum field is not  the preferred \emph{Bunch-Davies vacuum} state.  Nevertheless, in the collapse proposal, the quantum state of the field {\it after} the collapse is $|\Theta \r \neq |0 \ket$ (the analysis of a particular characterization of the post-collapse state has been done in Ref. \cite{11}). Therefore, the curvature perturbation responsible for the temperature anisotropies in the CMB is due to the expectation values $\bra \hat{y}_{\nk} \ket_\Theta$ and $\bra \hat{\pi}_{\nk} \ket_\Theta$, which, in principle, could generate detectable non-Gaussianities. These quantities are never considered in the standard accounts, and  it is clear that a further exploration of these ideas would   be required for  a serious  assessment of their value.

The other delicate issue related to the statistical aspects of the traditional approach is related to the ergodicity assumption.
 As we already saw in Sec. \ref{stand2},   the CMB bispectrum  was  defined as the three-point correlator of the $\alpha_{lm}$
    through $B_{m_1 m_2 m_3}^{l_1 l_2 l_3} \equiv \overline{\alpha_{l_1 m_1} \alpha_{l_2 m_2} \alpha_{l_3 m_3}}$.
The standard picture forces us  to  deal with the issue that  the rhs   represents  an  average over an ensemble of Universes, while we have
 but one realization  $\{ \alpha_{l_1 m_1}, \alpha_{l_2 m_2}, \ldots, \alpha_{l_n m_n} \}$. To overcome this issue, the standard approach
  relies on  an  ergodicity assumption,  which  identifies the average value of a certain quantity in  a process measured over time with the average value measured over the ensemble.

There are various issues  that  lead  one to be concerned  about this  assumption and the application to the situation at hand.
The  first thing we must be  aware of   is that ergodicity  is   a property of systems in equilibrium, and  it  is rather unclear why  this
should be valid regarding   the conditions   associated  with  the inflationary regime.

   Next,   as  already  mentioned,  the ergodicity assumption is translated, in the case at  hand,  into the  notion that
   the volume average of the fluctuations  behaves like the ensemble average {\it  ``the Universe may contain
     regions where the fluctuation is atypical, but with high probability most regions contain
   fluctuations with root-mean-square amplitude  close to $\sigma$},''
  and, thus,  one argues that  the probability distribution on the ensemble, translates to a probability distribution
  on smoothed regions of a determined size within our own Universe  \cite{44}.

  There are at least  three issues  that arise here:
\begin{itemize}

\item[i)] How  do we  go from the  arguments  supporting  ergodicity in time averages to  the corresponding  arguments  for  spatial  averages?

\item[ii)]  Regarding the CMB,  we,  in fact,  do not have access to    the  spatial  sections   that would  allow  us to
    investigate  the space  averages. We only  have    access to the  particular intersection of our past light come  with the
      3D hypersurface  of  decoupling. That is,  to a two-sphere  that we  see  as the source of the CMB  photons that reach us  today: the surface of last  scattering.
      How  do  we go  from spatial  averages  to   averages over that  two-sphere?
      \footnote{We note,  in relation to this point, that   there  are  intrinsic problems  in considering  ergodicity  of processes   within a two  sphere  as  discussed in Ref. \cite{49}.}

\item[iii)]  Each one of the quantities of interest,  $\alpha_{lm}$, is itself  already a  weighted  average  over the CMB  two-sphere [with the  weight function  given
    by the  corresponding $Y_{lm}(\theta, \varphi)$].
    Therefore,  what  would  be the role of  a new  average  over   the $m$'s ? Why do we  need to  perform any additional average?
    In other words, if one  is willing to accept that  the ensemble  averages  should  coincide  with  averages over the  two-sphere,  why  would one not  also
    accept that  the  weighted  averages  over the two-sphere   should coincide  with the equally  weighted   average over ensembles? If  we were to accept this, we  would conclude that the weighted  average  (with  weight  $Y_{lm}$ and  fixed $l$ and $m$) of $\delta T/T$ over the surface of  last  scattering for our Universe  should  coincide  with  the corresponding  weighted  average   of  that  quantity over the ensemble of  universes, without  any further averaging over $m$.
     The problem is  that the  latter would  be  zero,   but the former is  just  $\alpha_{lm}$,  which, empirically,  is   not  zero.
      Thus,  there must be something wrong  with our  arguments  and  assumptions.  One  should then consider what it is, and  why.

\end{itemize}

Let us  leave that rhetorical  question based  on a  position we are  rejecting and  consider again  the issue of  averaging over $m$.
     It seems clear  that what we   are  dealing  with here  are orientation averages:   The  different $\alpha_{lm}$  would   mix  among themselves  if we  were to redefine  the  orientation
      of the  coordinate chart   used to describe the  CMB two-sphere.  Thus, when we  look at  the  averages that are   actually  performed
      in connection with the   study of the primordial spectrum, we  see these are indeed   orientation averages.
 For instance,  the observational  quantity  $ C_l^{\textrm{obs}} = \frac{1}{2l+1}    \sum_m  |\alpha_{l m}|^2
$ is just the   orientation  average  value of the  magnitude of the $\alpha_{lm}$s  for a fixed   value of $l$. In the same way, we see that the angle-averaged bispectrum $B_{l_1, l_2, l_3}$ \eqref{avg}
is an orientation average for fixed $l$s and, as for the same reason as the one-point function, it is quite unclear how to identify orientation averages with ensemble averages. Thus, the  statistical analysis would be more transparent  if one would focus on the distribution of the quantities $B_{m_1 m_2 m_3}^{l_1 l_2 l_3} $.

As  we  saw, it is customary to take as an  estimator for the nonlinearity parameter the quantity  $ \hat{f}_\textrm{NL}$ defined  in Eq. \eqref{NL}.
This  seems  a  bit  problematic,  as   it  involves a mixture of theoretical   and   observational quantities. Ideally,
  one would  like to have   the  two  aspects rather  well separated.
 In  fact, even within the    standard approach,   for  the case of   the two-point functions,  we  have   on  one  hand
  the  theoretical  quantity,
  \beq
C_l^{\textrm{th}} = \frac{2}{\pi} \int k^2 P_\Psi(k) \Delta^2 (k) j_l^2 (kR_D) dk,
\eeq
and on the other hand the  observational quantity,

  \beq
C_l^{\textrm{obs}} = \frac{1}{2l+1}    \sum_m  |\alpha_{l m}|^2.
\eeq

This  independence of the definitions allows  one to  cleanly compare theory and   observation.
It, thus,  seems that  one  would  want to consider studying   the   aspects tied  to non-Gaussianity
 using a quantity that can be   equally
susceptible to  theoretical and  observational  determination.
Here, we  would like to propose,  based on the  considerations we  have  been discussing,  the  option we  present below.

First,   motivated   by  the  quantity   defined  in  Eq. \eqref{avg}, let us introduce  the definition of the  observed  bispectrum  as  the   orientation average

\beq\label{Ori-avg}
{\cal B}_{l_1 l_2 l_3}^{\textrm{obs}} \equiv \sum_{m_i}  \left( \begin{array}{lcr}
      l_1& l_2 & l_3  \\
     m_1 & m_2 & m_3
    \end{array}
    \right) ({\alpha_{l_1 m_1} \alpha_{l_2 m_2} \alpha_{l_3 m_3}})_{\textrm{obs}},
\eeq
and the definition of the  normalized   observational  reduced bispectrum  as the quantity
\bea\label{Norm-avgred}
{\tilde{b}}_{l_1 l_2 l_3 }^{\textrm{obs}} &\equiv& \left[ \sqrt{\frac{(2l_1+1)(2l_2+1)(2l_3+1) }{ 4\pi } } \left( \begin{array}{lcr}
      l_1& l_2 & l_3  \\
     0&0 & 0
    \end{array}
    \right) \right]^{-1} \nonumber \\ 
    &\times& {\cal B}_{l_1 l_2 l_3}^{\textrm{obs}},
\eea
 and, finally,  let  us define  the  magnitude  of the bispectral  fluctuations as

  \bea\label{Fluc}
       \mathcal{F}^{\textrm{obs}}_{l_1 l_2 l_3} &\equiv&
        \frac{1}{( 2l_1 +1)( 2l_2 +1)( 2l_3 +1) } \nonumber \\
        &\times& \sum_{m_i} |({\alpha_{l_1 m_1} \alpha_{l_2 m_2} \alpha_{l_3 m_3}})^{\textrm{obs}} - \mathcal{G}_{l_1 l_2 l_3}^{m_1 m_2 m_3}{\tilde{b}}_{l_1 l_2 l_3 }^{\textrm{obs}} |^2 . \nonumber \\
 \eea

One can then  compare this  pure  observational  quantity  with  the corresponding theoretical estimation  characterizing a suitable
 ensemble average,  where  each element of the  ensemble  is specified  by a
concrete choice of the  random numbers $x_{\vec k}$ that   we  have used  to represent the collapses.  That is,  one can  carry out  a
Monte Carlo simulation leading to  an ensemble of  possible  CMB   skies  characterized  by possible choices of   $x_{\vec k}$s,  and then
  characterize  each  one of   those  in terms of the corresponding value of $   \mathcal{F}^{\textrm{obs}}_{l_1 l_2 l_3} $. Finally one would   analyze
  the  degree to  which our  own  real  sky is   generic when characterized in that manner.  It seems  clear that   this  kind  of  theoretical calculation  or simulation cannot be  carried out in the
 standard approach,  as there is  no place there  for the concrete  randomness (characterized, in our approach, by the
 numbers  $x_{\vec k}$),  which   would be produced  in a simulation of our collapse proposal.

 Thus, the  study of the quantity displayed in Eq. (\ref{Fluc}) seems to offer  an approach to study  the issue at hand that  indeed has  the advantage  of allowing a direct   comparison  between  the   purely observational  quantities,  untainted  by theoretical  models,  and  the  quantities  that
are purely defined in terms  of such theoretical analysis.  This, in fact,  seems  to share  some of the spirit  of the analyses made  in Refs. \cite{30,31},  although  our proposal    provides a clear  option to compute   the observational  and  theoretical  quantities  in  complete separation, and  that   seems  not  to be  available in  the former.
  The  reason for this  seems   easy to understand:  The   fact that  we   maintain  a clear  distinction between  ensemble averages  and orientation averages  avoids  the possibility of the confusion associated    with the simple observation that   the  ensemble  average of  the quantity
\beq\label{Fl-2}
 ({\alpha_{l_1 m_1} \alpha_{l_2 m_2} \alpha_{l_3 m_3}})^{\textrm{obs}}- \mathcal{G}_{l_1 l_2 l_3}^{m_1 m_2 m_3}{\tilde{b}}_{l_1 l_2 l_3 }^{\textrm{obs}},
  \eeq
  appearing in  Eq. \eqref{Fluc},  vanishes identically.

 The  detailed analysis of  estimators like this    will be  carried out in  future works,  but   we  wanted to present it  as  an
  example  of the  type of studies  that could   be  motivated by our approach to  the  whole  question of the emergence of  structure from  quantum fluctuations in the inflationary early  Universe.

\section{Predictions and discussion}
\label{disc}

Focusing on trying to understand the essence of the
emergence of inhomogeneous and anisotropic features
from a quantum state, that is, homogeneous and isotropic
and in the absence of a measurement process,\footnote{In the early Universe, there were no observers or measuring
devices, and, in fact, the conditions for their emergence is the
result of the breakdown of such symmetries, so it would seem
very odd if one takes a view that they are part of the cause of that
breakdown.} has led us
to consider modifying the standard approach through the
incorporation of the collapse hypothesis.

We have seen in previous works that, despite the fact that
the motivation for such considerations seems to be purely
philosophical and tied to issues like the measurement
problem in quantum mechanics, the analysis has led us to
expect certain departures that could potentially be of observational significance.

In previous works, we have focused on two main observationally related issues: the shape of the spectrum and
the question of tensor modes. We have argued previously
that it would be very unlikely that one could find a scheme
in which the function $C(k)$ would be exactly a constant and
that some dependence on $k$ is likely to remain in any
reasonable collapse scheme, simply because we do not
expect those collapses to follow exactly the $\eta_k^c = Z/k$
rule for the time of collapse for each mode. Any remaining
dependency of $C(k)$ on $k$ will lead to slight deviations from
the standard form of the predicted spectrum. In fact, analyses of this issue have been carried out in Refs. \cite{8,9},
confirming these expectations. These have been used to
set the first bounds emerging from the CMB observational
data on these kinds of theories.

We have also stated, in earlier works, that the most clear
prediction of the novel paradigm we have been proposing
is the absence of tensor modes, or at least their very strong
suppression. The reason for this can be understood by
considering the semiclassical version of Einstein's equations and its role in describing the manner in which the
inhomogeneities and anisotropies arise in the metric. As
we have explained in our approach, the metric is taken to
be an effective description of the gravitational DOF, in the
classical regime, and not as the fundamental DOF susceptible to be described at the quantum level. It is, thus, the
matter degrees of freedom (which in the present context are
represented by the inflaton field), the ones that are described quantum mechanically and which, as a result of a
fundamental aspect of gravitation at the quantum level,
undergo effective quantum collapse (the reader should
recall that our point of view is that gravitation at the
quantum level will be drastically different from standard
quantum theories, and, in particular, it will not involve
universal unitary evolution). This collapse of the quantum
state of the inflaton field leads to a nontrivial value for
$\bra \hat{T}_{\mu \nu} \ket$,
 which then generates the metric fluctuations. The
point is that the energy momentum tensor contains linear
and quadratic terms in the expectation values of the quantum matter field fluctuations, which are the source terms
determining the geometric perturbations. In the case of the
scalar perturbations, we have first-order contributions proportional to $\dot{\phi}_0 \bra \hat{\dot{\delta \phi}} \ket $ while no similar first-order terms
appear as source of the tensor perturbations (i.e., of the
gravitational waves). Of course, it is possible that the
collapse scheme works at the level of the simultaneously
quantized matter and metric fluctuations, as has been presented in Ref. \cite{50}, although, as explained there, we would
find it much harder to reconcile that with the broad general picture that underlies our current understanding of physical
theories.

In the present work, we have focused on the modified
statistical considerations associated with this novel paradigm. We have argued that the collapse process itself could
be the source of non-Gaussian features. We discussed some
difficulties associated with the usual identification of measuring quantities with the quantum $n$-point functions and
particularly found that extending the standard arguments to
the one-point functions lead to disastrous disagreements
with observations.

We  have  shown that  our approach   provides  expressions   which have no parallel in the standard  formulations  and  which allow  a precise  identification of
the   location of the randomness, as  exemplified  by  our theoretical   formula \eqref{almcol} for  $\alpha_{lm}$ in terms of the random  numbers    characterizing  the collapses,  namely,  the
quantities
$x^{R}_{\vec k; 1}, x^{I}_{\vec k; 1}, x^{R}_{\vec k; 2}, x^{I}_{\vec k; 2}$.  This  kind of  expression facilitates  all resulting  statistical considerations,  and, in particular, it  is  the  basis for the theoretical estimation of  the quantity   \eqref{Fluc}.

  We   have  proposed  various   novel  ways   to look into the statistical aspects of the problem:

\begin{itemize}

\item[i)]  We  indicate the importance of exploring the true nature   of  the  one-point function    by studying the  degree  of deviation from  zero  of  the  complex quantity
   $\overline  \alpha^{obs} _l = \frac{1}{2l+1} \sum_l \alpha^{obs}_{lm}$ (i.e.  expanding  and refining the analysis of Ref. \cite{29})

\item[ii)]  We  have argued that it  is  worthwhile to study the  specific  form of the distribution of
  the  values of   the observed  quantities $|\alpha^{obs}_{lm}|$  for  each fixed  $l$.

\item[iii)]  We have  proposed   new  characterizations of the quantities normally  associated  with the bispectrum and the quantum  three-point functions  which can be   computed both in  purely  theoretically and  in a completely  observational fashion.   This is the quantity defined  in  Eq. \eqref{Fluc}.
\end{itemize}

 It is  clear that this work represents  only the  first step in the    study of the statistical aspects of the
  cosmic  structure and  its  generating  process
  during inflation,  within the context of the new paradigm  which  centers on the  collapse hypothesis.  Much more work remains
    to  be  done, but we hope this can become a research avenue of  great  richness and one
 that would lead to important insights,   with possible implications not only for the  generation of  structure itself but for the   modification of  quantum theory,  which  would underlie the  collapse   mechanism  and  which,  as  has   been   argued  before,  might  have   a deeper  origin at the  quantum/gravity  interface  \cite{5,16,17,51}.

\acknowledgments

The work of G. L. and D. S. is supported by CONACyT
Grant No. 101712 and PAPIIT Project No. IN107412-3.
G. L. acknowledges financial support by a CONACyT
postdoctoral grant. D. S. was supported in part by sabbatical fellowships from CONACyT and DGAPA-UNAM and the hospitality of the IAFE. The work of S. J. L. is supported by PIP N 0152/10 from Consejo Nacional de Investigaciones Cient\'ificas  y T\'ecnicas,  Argentina.

\appendix

\section{The Bunch-Davies vacuum is homogeneous and isotropic, correlations not withstanding.}\label{BD}

  {\bf Theorem: }The Bunch-Davies  vacuum state  (this, by the way, is also valid  for the Minkowski  vacuum state) is homogeneous and  isotropic.

   {\bf Proof:}

  The vacuum  state is defined  by
  {$\hat{a}_{\vec{k}} |0 \rangle =0$}.  This  is  supposed to  represent the  state of the  quantum field  after  few
      e-folds of inflation (up  to negligible  corrections  of order {$e^{-N}$},  with $N$ as the number of  e-folds),
  i.e. the exponential  expansion  of the Universe takes  the  metric and all fields to a very  simple state which, in particular, is
 highly symmetric. It  is  easy to  see  that the  state  $ |0 \rangle $ is H\&I. The   generator of spatial  translations  is $ \hat {\vec P} =  \sum_{\vec{k}}   \vec{k}  \ \hat{a}^{\dagger}_{\vec{k}} \ \hat{a}_{\vec{k}} $. So a translation   by  $\vec D$  leaves the   state unchanged, $e^{i \vec {D} \cdot \hat {\vec P} }   |0 \rangle=  |0 \rangle  $, and,  thus,  the state is  homogeneous. One can  equally  check that it is   isotropic  considering the    behavior of the state under rotations.
 Q.E.D.

  Furthermore, this  is  clearly not in  contradiction  with the existence of quantum correlations,    as they not imply a breakdown of the symmetry. This can  be   easily seen in a  Einstein-Podolsky-Rosen  setup.  Consider  a  state of two   spin-1/2 particles that result from the decay of  a spin 0 particle.

   Let us  assume that the decay occurring along the $ x $ axis (the particles  momenta  are $\vec P= \pm P \hat{x}$ with $\hat{x}$ the unit vector  in the  $ \vec x $ direction). Using the $\vec{z}$ polarization  states as a basis  for  the Hilbert  space of each  of the spin-1/2 particles,   the state of system after the decay  is

\beq
|\chi  \rangle = \frac{1}{\sqrt 2}( |+  \rangle^{(1)} |-  \rangle^{(2)}  + |+  \rangle^{(2)} |-  \rangle^{(1)} )
\eeq

The  state  can be  seen to be  invariant under rotations around the  $x$  axis (simply because  it is  an
  eigenstate  with  zero  angular  momentum along that axis). It is,  nevertheless, straightforward to compute the   correlations  between the operators
$ \hat S^{(1)} =\hat{\vec{\sigma}}^{(1)} \cdot  \hat{ \vec {N}}^{(1)} $  and $ \hat S^{(2)} =\hat{\vec{\sigma}}^{(2)} \cdot   \hat{\vec {N}}^{(2)} $
  corresponding to the projectors  of the spin  along the    vectors  (taken to  be   on the  $y-z $ plane) $ \hat{\vec {N}}^{(1)} $ and  $\hat{\vec {N}}^{(2 )} $, respectively.
  The result,   as is  well known  from the  studies of Bell's  inequalities, is proportional to  $\cos \theta$,   where $ \theta$  is the angle  between $\hat{ \vec {N}}^{(1)} $ and  $\hat{ \vec {N}}^{(2 )} $. Thus
  the  existence of these correlations  is in  no  conflict, whatsoever,
   with the  rotational invariance of the    state $ |\chi  \rangle $. It seems that   the  belief that  there  is   something in the correlations that  indicates    the breakdown of the symmetry is tied to  some  implicit intuition  suggesting that  each one   of   the particles  is  in   a definite  state of spin, even  before  there are any measurements involved. We,  of course, know that such  notions  are in strong  conflict  with both  quantum theory and the  experimental facts.

\section{Divergence of $\langle 0 | \hat{\Psi}^2 (\vec{x},\eta) | 0 \rangle$} \label{div}

In order to illustrate a common misconception about the quantity $\langle 0 | \hat{\Psi}^2 (\vec{x},\eta) | 0 \rangle$, let us consider the following argument: The gravitational potential that gives temperature anisotropy is not $\hat \Psi (\nk)$ from the primordial era but $\hat{\Psi} (\nk) \Delta(k)$, where $\Delta(k)$ is the transfer function. Since $\Delta(k) \propto \ln (k)/k^2$ for large $k$, $\langle 0 | \hat{\Psi}^2 |0 \rangle$ is convergent in the UV regime.

The previous statement is not correct.   The   transfer functions characterize the physics  that is  relevant  after  the  end  of inflation (i.e.  the physics  characterizing the behavior of the   radiation and particles  that  emerge  as  the result of  reheating,  including, for instance, the  famous  plasma oscillations). That is why they are   called  transfer functions. They indicate  how  the perturbations  that were present   during inflation (to  be exact  at  its  end point)  evolve during the  radiation-dominated  era  into the   perturbations that  are  present   at the time of decoupling,  which is   the relevant one for what  we  see in the CMB.
  The transfer functions   are, of course, not  relevant  at all   during the  inflationary era itself, which is  the   era we  are focusing on (and   the one  in which  we  argue the collapse should  occur). The issue, related to the divergence of $\langle 0 | \hat{\Psi}^2 (\vec{x},\eta) | 0 \rangle$, clearly refers  to  the inflationary era. The  (rhetorical) question  we are posing is the following: If  we  do not take   the  expectation value of $\hat \Psi_k$ to  be the inflationary   prediction  for  the $ \Psi_k$,  as that  would  be  zero,  and  we are instead  instructed to  compute  the vacuum expectation value for   quantity $\hat \Psi_k \hat \Psi_{k} $  and to use it  in order to  make  our estimates  of $ \Psi_k$,  then,  why  would it be incorrect to  compute the  vacuum  expectation value of $\hat \Psi (x) \hat \Psi (x) $  and take it  as  an estimate of
   the   value of $ \Psi (x)$ (during the inflationary   era)? The  issue  is  that such an estimate would be   infinite,  and  then  the  whole
    scheme of perturbation theory on which the   treatment is  based  would be invalid. We would, therefore, not be  able to rely on it, either to consider the  study of the
   predictions regarding the  radiation-dominated  era  or the  CMB.

\section{ On the Interpretation of Quantum Theory and the  Cosmological  Context}\label{QT}

Here, we will briefly consider, for the convenience of the
reader, some of the most common views we have found
among colleagues regarding the interpretation of quantum
physics and their implications for its application to the
cosmological problem at hand, and, at the same time, we
present our basic perspective on such views. A more detailed discussion of these issues has been presented in
Ref. \cite{5}, and the reader is advised to turn to that reference
for a thorough analysis of the alternative postures taken by
researchers in the field.

The issue we are facing is, of course, related to the so-called measurement problem in quantum mechanics \cite{52}.
Any reasonably complete discussion of this question and
the broader one concerning the interpretation of quantum
mechanics would require much more space than what can
reasonably be accommodated here, so we point the reader
to some of the literature \cite{53}. Here, we will merely present
our version of the status of the general problem, touching,
when appropriate, on the particular instance that concerns
us here: the cosmological setting. That issue has not received to much attention from the physics community, with
notable exceptions represented by Penrose \cite{17}, Hartle \cite{7},
and others.

\begin{itemize}

  \item { (i) } {\it Quantum physics  as  a theory of  statistical physics}--A point of view indicating that quantum mechanics acquires meaning only as it is applied to an
ensemble of identically prepared systems \cite{54}.
Thus according to this view, a single atom, in
isolation, is not described by quantum mechanics.
We must avoid getting confused by the correct, but
simply distracting, argument that atoms in isolation do not exist. The issue is whether, to the
extent to which one can neglect its interactions
with other systems, quantum mechanics is applicable to the description of a single atom. One
might wonder about the meaning of that question,
given that we cannot say anything about the atom
without making it interact with a measuring device. The question is simply whether or not applying the formalism of quantum mechanics to treat the single isolated atom can be expected to yield
correct results regarding subsequent measurements. It is sometimes argued that this is a nonsensical question, as these results are always statistical in nature. However, in fact, this statement is not accurate; for instance, if the atom (e.g.,
of hydrogen) was known to have been prepared in
its ground state, the probability of detecting it in
some other energy eigentstate is zero. Thus, there
must be something empirical in the description of
that single atom by its usual quantum mechanical
state. The notion that quantum mechanics is not
applicable to a single system \cite{55} is, thus, simply
incorrect. However, the most important point in
relation with the issue that concerns us in this
article is that taking a posture like this about
quantum physics, would be admitting from the
beginning that we would have no justification in
employing such a theory in addressing questions
concerning the unique Universe. Note that the
situation would not be helped if we assumed that
there exists, in some sense of the word, an ensemble of universes, as we would, in principle,
have access just to one: ours. Advocates of the
standard accounts of inflation usually invoke some sort of identification of the statistics with an ensemble of universes and the statistics within one
single Universe. However, as we have argued
throughout this work, it is paramount to avoid
confusion between those types of statistical measures as a matter of principle, even if one would
later want to argue they might be identified in
some special cases. It should be clear that in order
to be able to argue for any such identification, one
must be in a position to say something about the
individual system. In fact, we can imagine considering any individual system whatsoever, say, a cloud of
gas, and constructing an ensemble of similar systems
by performing say ``all possible rotations and translations of the system.'' It is clear that the resulting
ensemble is, by construction, homogeneous and isotropic. Now, can this be used to say anything about the
original system? Clearly, it cannot, simply because our
starting point was a completely arbitrary system. Thus,
if we negate from the start that our theory could say
anything about an individual system, there is no way
we can apply it to our Universe. Furthermore, going
back to the general case, if a quantum state serves only
to represent an ensemble, how is each element of the
ensemble to be described? Perhaps, it cannot be described at all. Then, how are we supposed to make statistics over the attributes of such systems?

\item{ (ii) } {\it Quantum physics as a theory of human knowledge}--According to this view, the state of a quantum system is not considered as reflecting anything
``objective'' about the system in question but just
provides a characterization of ``what we know about
the system''\footnote{One can find statements in this sense in well-known books,
for instance, ``quantum theory is not a theory about reality, it is a
prescription for making the best possible predictions about the
future, if we have certain information about the past'' \cite{56}. See
also Ref. \cite{57}.}. This attitude, naturally rises the
question of what there is to be known about the
system if not something that pertains to the system.
Advocates of this point of view often answer in
terms of correlations between the system and the
measuring devices. This leads us to consider the
question of the significance of these correlations.
Note that the meaning of the word ``correlation''
implies some coincidence of certain conditions pertaining to one object, with some other conditions
pertaining to a second object. Therefore, if a quantum state describes such a correlation, there must be
some meaning to the conditions pertaining to each
one of the objects: the quantum state and the system.
Are not these, then, the very same conditions that
are described by the quantum mechanical state and
those that correspond to the object? If the answer
is ``no'', it must mean that there are further
descriptions of the object that cannot be cast in the
quantum mechanical state vector. On the other hand,
if the answer is ``yes'', we would be implying that
the state vector says something about the object in
itself. Independently of these issues, it seems rather
clear that if we follow the above described view, we
would have abandoned the possibility to consider
questions about the evolution of the Universe in the
absence of sapient beings or to consider the emergence, in that Universe, of the conditions that are
necessary for the eventual evolution of humans,
while making use of our quantum theory.

\item{ (iii)} {\it Quantum physics as a noncompletable description of the world}--Within this class we consider any
posture that effectively, if not explicitly, states,
``The theory is incomplete, and no complete theory
containing it exists or will ever do.'' Such a view
includes any posture indicating we should use
quantum mechanics ``as we all know how'' and
supported by the observation that no violation of
quantum mechanics has ever been observed.\footnote{
In practice, this view is essentially indistinguishable from the
so-called \emph{for all practical purposes} approach to the matter \cite{58}.} At
this point we must note that although this is a
literally correct statement, the prescription refers,
in fact, to the Copenhagen interpretation, which, as discussed above raises severe interpretational issues that become insurmountable once we leave
the laboratory and attempt to apply quantum theory
to something like the Universe itself.
The fact is that, in situations in which one cannot
identify the system and the measuring apparatus, the
observables that are to be measured, the entity
carrying out those measurements, and the time at
which the measurements take place, the theory does
not provide any clearly defined scheme specifying
how to make the desired predictions. Thus applies,
in particular, to the questions pertaining to the early
Universe. However, according to such pragmatic
approach we should be satisfied with the fact that
the predictions have, in fact, been made and that
they do seem to agree with observations. The problem is, that in the absence of a well-defined set of
rules, it becomes quite unclear whether or not such
``predictions'' follow or not from the theory without
the use of extraneous and ad hoc, but convenient
hypothesis suitably introduced in connection to the
particular problem at hand. Especially suspicious
are, of course, those predictions that are, in fact,
retrodictions, and, on this point, we should recall
that, long before inflation was invented, Harrison
and Zel'dovich \cite{59} had already concluded what
should be the form of the primordial spectrum,
based on quite general observations about the nature
of the large-scale structure of our Universe.

\item{ (iv) } \emph{Quantum physics as part of a more complete
description of the world}--Completing the theory
would require something that removes the need for
invoking any sort of a priori distinct notions of
external measurement apparatus, an external observer, etc. One proposal of this kind is provided
by D. Bohm's ``pilot wave theory'' \cite{60}, and, in
particular, we note a specific proposal to apply
such ideas to the cosmological problem at hand
\cite{61,62}. As we have mentioned, there are other
proposals invoking something like the dynamical
reduction models proposed by Pearle \cite{23} and/or
Ghirardi \textit{et al}. \cite{18} and the ideas of R. Penrose
about the role of gravitation in modifying quantum
mechanics in the merging of the two aspects
of nature \cite{17} (see also Ref. \cite{63}). In the
context of inflationary cosmology, the works
(Refs. \cite{4,8,9,10,14}) are an example in which the
issues are faced directly. Those works represent
the position we favor, and which is inspired, in
part, by the arguments made in Refs. \cite{17,18,20}.

\item{ (v) } \emph{Quantum physics as a complete description of the
world}--Here we refer to any of the postures indicating that quantum mechanics faces no open issues
and that, in particular, the measurement problem has
been solved. The advocates of this position fall into
groups identified with one of the the main currents:
those that subscribe to ideas along the so-called
``many world interpretation of quantum mechanics,'' and consider this to be a solution to the measurement problem, and those that consider that this
problem has been solved by the various considerations involving ``decoherence.'' We consider that the
many world interpretation does very little to ameliorate the measurement problem, as there is a mapping between what in that approach would be called
the ``splittings of worlds''\footnote{It is often claimed that there is no splitting of the worlds in the
many worlds interpretation, but the fact of the matter is that,
whenever people make use of it, they cannot avoid talking about
things such as ``our branch,'' ``the realms we perceive,'' or other
notions that implicitly make use of a notion that is essentially just that of a splitting of the world. One can see this in each specific application of the idea, by focusing on the complete description of what one would take as ``the relevant state describing our
reality'' and following it in time backward and forward. In the
inflationary situation at hand, this is easily done by focusing on
the symmetry of the state describing the quantum fields.
} and what would be
called measurements in the Copenhagen interpretation. In fact one can see that almost every issue that
can be raised against in the context of the latter has a
corresponding one in the many worlds interpretation
choice of basis or context in dealing with measurement problems correspond to the selection of basis
for the ``world splittings,'' time of such splittings
would need to include those that one takes as the ``times of measurements,'' and so forth. In other
words, concerning a specific measurement situation,
and the corresponding description within the Many
Wolds Interpretation, the issues would be the following: When does a world splitting occur? Why, and
under what circumstances does it occur? What constitutes a trigger? And, finally, what selects the basis
in which such splittings takes place? The ideas based
on a decoherence type solution and its shortcomings
will be discussed in some detail in Appendix \ref{Problems}.

\end{itemize}

\section{Shortcomings of the usual explanations  of the emergence of  primordial  inhomogeneities  and anisotropies.}\label{Problems}

We offer here a very brief version of the discussion
presented in Ref. \cite{5} of why we find the most widely
held views on the way of addressing our problem as
lacking. In our experience, these are the ``decoherence
approach'' (perhaps supplemented by the many worlds
interpretation of quantum theory) and the ``consistent or
decohering histories approach.''

We will, thus, offer some considerations regarding the
degree to which these two proposals do, in general, offer a
``solution'' to the measurement problem and, particularly,
of their applicability in the present context. Again, we
suggest turning to Ref. \cite{5} for a more exhaustive discussion
of all these issues.
   
    \subsection{Decoherence}

Decoherence is a direct prediction of quantum mechanics, with very important implications in many experimental situations. The central observation is that, in the
general experimental setting involving a quantum mechanical system, one should take into account the fact
that generally the system becomes entangled with the
environment, consisting of degrees of freedom that are
not under the control of the experimentalist and which
are, moreover, uninteresting from the point of view of
what one is interested in measuring. On the other hand
many colleagues seem to think that it has implications
that go well beyond that and which represents a complete
and satisfactory solution of the measurement problem in
quantum mechanics. This is not the case, and the interested reader is directed to consult the literature on the
matter (see, for instance, Ref. \cite{64}).
Here, we will limit ourselves to indicating the postures
that, in this regard, are held by several people that have
considered the issue at length, in order to contrast them with
the prevalent notions among inflationary cosmologists.
Let us start with the explicit conclusion by
A. Neumaier \cite{65}:

    \begin{quotation}
     ``Many physicists nowadays think that decoherence
provides a fully satisfying answer to the measurement problem. But this is an illusion.''
    \end{quotation}

   Also worthwhile is the warning by M. Schlosshauer \cite{66} against misinterpretations:
   
   \begin{quotation}

``...note that the formal identification of the reduced
density matrix with a mixed state density matrix is
easily misinterpreted as implying that the state of
the system can be viewed as mixed too.... the total
composite system is still described by a superposition, it follows from the rules of quantum mechanics that no individual definite state can be attributed
to one of (the parts) of the system...''

\end{quotation}

and the explicit refutation by E. Joos \cite{67}:

\begin{quotation}

``Does decoherence solve the measurement problem? Clearly not. What decoherence tells us is that
certain objects appear classical when observed, but
what is an observation? At some stage we still have
to apply the usual probability rules of Quantum
Theory.''
\end{quotation}

Thus, the decoherence ideas, even if taken together with
the many worlds interpretation, clearly fail to offer a
satisfactory resolution of the matter in general \cite{68}, and,
in particular, it fails to do so in connection for the situation
we face here.

Let us end by noting that even W. Zurek, one of the most
well-known researchers in the field of decoherance, states
unequivocally that \cite{69}:

\begin{quotation}
 ``The interpretation based on the ideas of decoherence and ein-selection has not really been spelled out
to date in any detail. I have made a few half-hearted attempts in this direction, but, frankly, I was
hoping to postpone this task, since the ultimate questions tend to involve such ``anthropic'' attributes of the ``observership'' as ``perception'', ``awareness'', or ``consciousness'', which, at present, cannot be modeled
with a desirable degree of rigor.''
\end{quotation}

The point is that in the context of inflationary cosmology, in which we want to explain the emergence of the
seeds of cosmic structure, and, thus, the emergence of the
conditions that would eventually create the conditions for
the emergence of galaxies, stars, biology, and intelligent
life, we cannot even hope to rely on any of those anthropic
notions. Thus, decoherence does not represent an adequate
solution to the problem at hand.

   \subsection{Consistent Histories}

   The general scheme as described in \cite{70}, is based  on  the   consideration, given  a quantum state  of the system
   $ | \Phi  \rangle$, or, more  generally, a density matrix $\hat \rho$,  for the system at time $t_0$,   of  families of histories   characterized  by  a set  of  projection operators $\lbrace \hat P_{n} (t_n)\rbrace $,  each of  which   is
   associated with the system possessing a value of certain physical property  in  a given  range   at a given time.\footnote{In  the  cosmological  setting, one  must use  a  subtler  relational time  approach \cite{7},
   in which  one of the dynamical variables is used   as an effective time  parameter.   The cosmological  scale factor  is a popular choice.}  That is,  each  one of the projector operators  is  associated  with a  certain range  within the
   spectrum of  a given observable.    A given family  $F $ of such projectors, is  called  self-consistent if the resulting histories  do not interfere  among themselves. In that   case,  one  may  consistently  assign  probabilities  to  each    individual  ``coarse-grained history"  within the family.\footnote{The   characterization of the histories  as  coarse-grained is  meant to reflect the fact that the projection operators $ \hat P_{n} (t_n) $   are generically associated  with a finite range of eigenvalues  rather than a single  eigenvalue.}

The  probability  assigned  to  one   a particular  coarse-grained   history  within  a consistent family  is given  by
   \bea\label{Prob}
   P &=& Tr ( \hat P_{n}(t_n) U(t_n, t_{n-1} ) \hat  P_{n-1} U(t_{n-1}, t_{n-2} )  ...... \hat P_{2}U(t_{2}, t_{1} ) \nonumber \\
   &\times& \hat P_{1}U(t_{1}, t_{0} ) \hat\rho  U(t_n, t_0 )^\dagger ),
   \eea
   where  $ U (t, t')$ stands for the  standard unitary evolution operators  connecting  the two  times.
   This approach, apparently,   has a good  number  of   followers within the  cosmology community,  but it has  been  subjected to  some strong criticisms  in the foundational  physics community \cite{71}.

The issue is that, although the scheme works fine once
one has selected a particular decoherent family $F$, there
exist, in principle, an infinitude of other such decoherent
families $F'$, which are, however, mutually inconsistent,(i.e., there are elements of $F$ and $F'$ that do interfere,
and, thus, $\lbrace F \rbrace\cup \lbrace F' \rbrace$ is not a decohering set of histories).
This problem, which is well known to the advocates of
this approach should according to them be addressed by
the ``single family rule,'' which indicates one should
never consider, simultaneously, more than one family.
Moreover, according to this approach, we should never
make any logical inferences while considering together
two inconsistent sets, as they can produce logical contradictions (see, for instance, the example discussed in
Sec. 3 of Ref. \cite{72}). As noted in Ref. \cite{73}, it is unclear
what would justify this rule within a reasonable ontological view of what the theory is describing.

The issue becomes then how should we select a
particular family to be that from which the particular
history that represents ``the actual one'' is to be chosen.
It seems very reasonable that the fact that one assigns
probabilities within a family indicates that the interpretation must be that one of the histories in that family is
``actualized'' in our world. Otherwise, one must wonder
what these probabilities refer to [i.e., the probabilities
assigned according to Eq. \eqref{Prob} are probabilities of
what? (see, however, Ref. \cite{72})]. Let us emphasize
once more that we do not want to take the view that they are probabilities of ``observing a certain value of
a physical quantity when that quantity is measured''
because the whole point of this program is to have an
interpretational framework for quantum theory that avoids
using concepts like measurements or ``observations'' in the
discussion (otherwise, one might as well have retained the
Copenhagen interpretation).

The fundamental problem is that there is in principle, no
clear way to single out one specific family without relying
on an \emph{a priori} given set of questions one interested on those associated with the quantities whose spectral characteristics one chooses to construct the family--and this ambiguity leads to serious interpretational difficulties \cite{73}.

In a specific situation, we might be guided in making the
``appropriate'' choice, by the questions the experimental
setup is ``asking'' (in fact, this has a close analogy with the
use of Bohr's rule in a given experiment). Nevertheless the
fact remains that, in general, without such an common sense,
or practical guidance, there is no well defined procedure
indication how to select the family. We must here emphasize
that one is not asking how to select a particular history within
the family but how to select a particular family of consistent
histories from within the collection of all possible decoherent
families. The problem here is: what justifies considering that
the experimental setup corresponds to asking a particular
question; this seems to implicitly assume that the measuring
apparatus is always in a state of definite value for the measured quantity and never in the superposition of states of that
type. This seems very close to what one does in adopting the
Copenhagen interpretation.

Returning to our specific problem, of describing the
evolution of the very early universe, there is simply no
recipe provided by the theory, that would dictate the selection of the appropriate projector operators and, thus, of
the appropriate family (if we require a description that does
not make use of the fact of our own existence and our own
asking of certain questions as part of the input).

Let us see a clear manifestation of this problem in the
cosmological situation of interest: Consider the family of
projector operators as is done in Ref. \cite{74}, and obtain their
results, but then note that, alternatively, we might consider
the family in what follows. We next define $P_{\textrm{HI}}$ to be the
projector into the intersection of the kernels of the generators of translations and rotations (note that it is the projector onto the space of homogeneous and isotropic states).
Let us further define $P_{\textrm{non}} \equiv  I - P_{\textrm{HI}}$ the orthogonal projector. Take the initial state for the quantum fluctuations
(usually called the vacuum) $| \Phi_0 \ket$, and note that it is homogeneous and isotropic.

The next step is to consider an arbitrary collection of
values for time ${ t_i }$ and construct the family associated with
that initial state and the two projector operators $P_{\textrm{HI}}$ and
$P_{\textrm{non}}$ at each of those times. One can easily check that this
procedure defines a family of consistent histories, simply
because the dynamics (characterized by the operators $U$)
preserves the symmetries (homogeneity and isotropy).

Consider now the question of what the probability is
that (at a given time, characterized in the appropriate
relational way) the Universe is isotropic and homogene-
ous. Evaluating this using the formula \eqref{Prob} (and starting
with the vacuum state), we find that any history containing the orthogonal projector at any time $P_{\textrm{non}}$ has zero
probability, while the history containing only the operators $P_{\textrm{HI}}$ has probability one. We seem to be led to conclude that, at any time, the Universe is homogeneous and
isotropic. It, thus, can have no inhomogeneities or anisotropies at all. We would then have to face not only such a
problematic prediction but also the fact that the approach
we followed has led us to two conflicting conclusions: this
latter one and the one obtained in, say, Ref. \cite{74}.

In fact, this problem is similar to those considered in
Ref. \cite{71} and that discussed in Sec. 3 of Ref. \cite{72}. The
posture advocated in Ref. \cite{72} is that one should accept all
the different families and use only the appropriate one in
connection with the particular question one is asking in
order to make ``bets about the future,'' while at the same
time renouncing the idea that there is a single objective
reality. As discussed in Ref. \cite{73}, such a posture seems
unsustainable in addressing the fact that we humans seem
to coincide regarding our appreciation of the ``world out
there.''

Apparently, if choosing to accept the consistent histories
approach to the quantum theory, in general, one would
have to adopt a rather problematic position close to that
of posture (ii) in Appendix \ref{QT}, with the difficulties already
mentioned there. It seems that this is not a satisfactory
situation regarding something that ought to serve as a
fundamental theory and, in particular, to help us deal
with the quantum aspects of the early Universe. The interested reader is referred to the literature, particularly, to
the works referred to above, for much more extensive
discussions on the matter.

\end{document}